\newif\ifarxiv
\arxivfalse

\ifarxiv
\documentclass[a4paper,9pt,abstract=true]{scrartcl}
\usepackage{times}
\else
\documentclass[preprint,9pt]{elsarticle}
\fi

\AtBeginDocument{%
	}
		
\usepackage{hyperref}
\usepackage{url}

\usepackage[english]{babel}
\usepackage{microtype}

\usepackage{hyperref}
\usepackage[T1]{fontenc}
\usepackage[scaled=0.81]{beramono}
\usepackage{multirow}
\usepackage{booktabs}
\usepackage{graphicx}
\usepackage{amssymb}
\usepackage{amsmath}
\usepackage[draft]{changes}

\setcitestyle{square,numbers,sort}

\usepackage{float}
\ifarxiv\else

\makeatletter
\let\@afterindenttrue\@afterindentfalse
\makeatother
\fi

\journal{Journal of Systems and Software}

\usepackage{booktabs}
\usepackage{listings}
\usepackage{lstautogobble}

\lstdefinestyle{displayed}{
  numbers=none, %
  firstnumber=last, 
  breaklines=true,
  breakatwhitespace=false,         
  keepspaces=true,                 
  numbersep=2pt,
  showspaces=false,                
  showstringspaces=false,
  showtabs=false,                  
  frame=tb,
  captionpos=b,                    
  numberstyle=\scriptsize,
  tabsize=2,
  captionpos=b,
  xleftmargin=0mm, %
  xrightmargin=0mm, %
  basicstyle=\ttfamily\small,
  keywordstyle=\bfseries\ttfamily,
  commentstyle=\color{darkgray}\itshape\ttfamily,
  keepspaces=true,
  columns=fixed,
  escapeinside={(*}{*)},
  mathescape=true,
  showstringspaces=false
}

\newcommand{\J}[1]{\mbox{\lstinline[basicstyle=\ttfamily,language=Java]|#1|}}

\usepackage{graphicx}

\usepackage{dsfont}
\usepackage{latexsym}
\usepackage{csquotes}
\usepackage{listings}
\usepackage{float}
\usepackage{multirow}
\usepackage[scaled]{helvet}
\usepackage[noend]{algpseudocode}
\usepackage{mathrsfs}
\usepackage[linesnumbered, ruled, lined]{algorithm2e}  %
\usepackage{mathpartir}
\usepackage{mathtools}
\usepackage{stmaryrd}
\usepackage{url}
\usepackage{textcomp} 
\usepackage{bbm}
\usepackage{verbdef}
\usepackage{xspace}
\usepackage{verbatim}
\usepackage{lipsum}
\usepackage{wrapfig}

\usepackage[shortlabels,inline]{enumitem}
\usepackage{pifont}
\usepackage{varwidth}
\usepackage{xpatch}
\usepackage{multirow}
\usepackage{xcolor}
\usepackage[export]{adjustbox}
\usepackage{caption}
\usepackage{subcaption}
\usepackage{setspace}
\usepackage[capitalize]{cleveref}
\usepackage{makecell}
\usepackage{dashrule}
\usepackage{comment}
\usepackage{arydshln}
\usepackage{ifsym}
\usepackage{mdframed}

\setlist[enumerate]{label=\emph{\roman*})}
\setlist[description]{font=\normalfont\bfseries,labelindent=0mm,style=multiline,leftmargin=8mm}

\usepackage{flushend}

\usepackage{tikz}
\usetikzlibrary{calc,positioning,math,arrows,shapes,fit,intersections}
\usepackage{pgfplots}
\pgfplotsset{compat=newest}

\usepackage{numprint,relsize}
\xspaceaddexceptions{\%}  %

\usepackage{xstring,xparse}

\usepackage{fontawesome}

\newcommand{\includeOK}[1][green]{\text{\color{#1}\faCheck}}
\newcommand{\includeNO}[1][red]{\text{\color{#1}\faClose}}

\usepackage{tcolorbox}

\DeclareDocumentCommand{\finding}{m}{%
  \begin{tcolorbox}[
    boxsep=0pt,
    left=6pt,
    right=6pt,
    top=2pt,
    bottom=2pt,
    colframe=white,
    colback=bcol!20!white,
    boxrule=0.3pt, %
    ]
    \small\textit{%
      #1%
      }
\end{tcolorbox}
}

\newif\ifdraft
\drafttrue

\newif\ifblind
\blindtrue

\newif\iflong
\longtrue

\iflong
\else
\makeatletter

\newcommand\niceparagraph{\@startsection{paragraph}{4}{\z@}%
  {0\p@ \@plus 1\p@ \@minus 1\p@}%
  {-0.5em \@plus -0.22em \@minus -0.1em}%
  {\bfseries\normalsize}}
\makeatother
\renewcommand{\paragraph}[1]{\niceparagraph{#1}}

\fi

\newcommand{\prevarank}{{\smaller[0.5]{\textsc{Preva\-Rank}}}\xspace}
\newcommand{\tool}{\prevarank}

\DeclareDocumentCommand{\freq}{o o}{%
  \ensuremath{%
    \mathds{F}%
    \IfNoValueTF{#1}{}{\IfStrEq{#1}{}{\!}{_{#1}}}%
    \IfNoValueTF{#2}{}{\left({#2}\right)}%
}}

\DeclareDocumentCommand{\bugc}{s o}{%
  \ensuremath{%
    \IfBooleanTF{#1}{\widehat{\mathsf{c}}}{\mathsf{c}}\IfNoValueTF{#2}{}{(#2)}%
  }}
\DeclareDocumentCommand{\patchk}{o}{%
  \ensuremath{\mathsf{k}\IfNoValueTF{#1}{}{(#1)}}%
}

\DeclareDocumentCommand{\passat}{m}{\ensuremath{\mathit{pass@}#1}}

\DeclareMathOperator*{\argmax}{argmax}

\definecolor{acol}{HTML}{d7191c}
\definecolor{bcol}{HTML}{fdae61}
\definecolor{ccol}{HTML}{ffffbf}
\definecolor{dcol}{HTML}{abd9e9}
\definecolor{ecol}{HTML}{2c7bb6}

\definecolor{fcol}{HTML}{a6611a}
\definecolor{gcol}{HTML}{dfc27d}
\definecolor{hcol}{HTML}{80cdc1}

\definecolor{rankcol}{HTML}{018571}

\newcommand{\reducedstrut}{%
  \vrule width 0pt height .9\ht\strutbox depth .9\dp\strutbox\relax%
}

\newcommand{\best}[2][bcol!50]{%
  \begingroup
  \setlength{\fboxsep}{0pt}%
  \colorbox{#1}{\reducedstrut#2\/}%
  \endgroup
}

\tikzset{
   declare function={gamma(\z)=
     2.506628274631*sqrt(1/\z)+ 0.20888568*(1/\z)^(1.5)+ 0.00870357*(1/\z)^(2.5)- (174.2106599*(1/\z)^(3.5))/25920- (715.6423511*(1/\z)^(4.5))/1244160)*exp((-ln(1/\z)-1)*\z;},
   declare function={gammapdf(\x,\k,\theta) = 1/(\theta^\k)*1/(gamma(\k))*\x^(\k-1)*exp(-\x/\theta);}
}

\DeclareDocumentCommand{\freqDistro}{s m O{!} O{2,2} O{9,0.5} O{1.1,1.6}}{%
  \resizebox{#2}{#3}{
    \begin{tikzpicture}
      \begin{axis}[
        samples=100,
        xlabel={\IfBooleanT{#1}{patch kind $k$}},
        ylabel={\IfBooleanT{#1}{$\freq[c][k]$}},
        axis lines=middle,
        axis line style={ultra thick, -},
        xlabel style={at={(rel axis cs:0.5,-0.04)}, anchor=north, align=center},
        ylabel style={at={(rel axis cs:0,0.8)}, anchor=east, align=center},
        xmin=0, xmax=6,
        xtick={0,0.5,...,10},
        xticklabels={},
        xtick style={very thick}, %
        major tick length=10pt,
        ytick=\empty,
        y axis line style={draw=none},
        legend style={at={(1,1)}, anchor=north east, draw=none,
          legend image code/.code={%
            \draw[##1,line width=4pt] plot coordinates {(0cm,0cm) (0.3cm,0cm)};
          },},
        height=6cm, width=6.4cm,
        ]
        \addplot [fill=acol, draw=acol, very thick, opacity=0.7, domain=0:6.0] {gammapdf(x,#4)} \closedcycle;
        \IfBooleanT{#1}{\addlegendentry{$c_2$}}
        \addplot [fill=bcol, draw=bcol, very thick, opacity=0.5, domain=0:6.0] {gammapdf(x,#5)} \closedcycle;
        \IfBooleanT{#1}{\addlegendentry{$c_3$}}
        \addplot [fill=dcol, draw=dcol, very thick, opacity=0.6, domain=0:6.0] {gammapdf(x,#6)} \closedcycle;
        \IfBooleanT{#1}{\addlegendentry{$c_1$}}
      \end{axis}
    \end{tikzpicture}%
  }%
}
\begin{document}
	\sloppy
\begin{frontmatter}

\title{\tool: Ranking Plausible Patches \\by Historic Feature Frequencies}

\ifarxiv
  \author{
    Shifat Sahariar Bhuiyan$^1$
    $\ \cdot\ $
    Abhishek Tiwari$^1$
    $\ \cdot\ $
    Yu Pei$^2$
    $\ \cdot\ $
    Carlo A. Furia$^1$
    \\[2mm]
    $^1$ USI Università della Svizzera italiana, Lugano, Switzerland
    \\
    $^2$ The Hong Kong Polytechnic University, Hong Kong, China
  }
\else
\author[usi]{Shifat Sahariar Bhuiyan}
\author[sdu]{Abhishek Tiwari\corref{cor1}}
\cortext[cor1]{Corresponding Author}
\author[hk]{Yu Pei}
\author[usi]{Carlo A.\ Furia}
\affiliation[usi]{organization={USI Università della Svizzera italiana},
	             city={Lugano},
	            country={Switzerland}}

\affiliation[sdu]{organization={University of Southern Denmark},
		city={Odense},
		country={Denmark}}

\affiliation[hk]{organization={The Hong Kong Polytechnic University},
	city={Hong Kong},
	country={China}}
\fi

\ifarxiv\maketitle\fi

\begin{abstract}
  Automated program repair (APR) techniques
  have achieved conspicuous progress,
  and are now capable of producing genuinely \emph{correct}
  fixes in scenarios that were well beyond their capabilities
  only a few years ago.
  Nevertheless, even when an APR technique can find a correct fix
  for a bug, it still runs the risk of \emph{ranking} the fix
  lower than other patches that are plausible
  (they pass all available tests) but incorrect.
  This can seriously hurt the technique's practical effectiveness,
  as the user will have to peruse a larger number of patches
  before finding the correct one.

  This paper presents \tool,
  a technique that ranks plausible patches produced by any APR technique
  according to their feature similarity
  with historic programmer-written fixes for similar bugs.
  \tool implements simple heuristics,
  which help make it scalable and applicable to any APR tool that
  produces plausible patches.
  In our experimental evaluation,
  after training \tool on the fix history of 81 open-source Java projects,
  we used it to rank patches produced by 8 Java APR tools on 168 Defects4J bugs.
  \tool consistently improved the ranking of correct fixes:
  for example, it ranked a correct fix within the top-3 positions
  in 27\% more cases than the original tools did.
  Other experimental results indicate that \tool works robustly
  with a variety of APR tools and bugs, with negligible overhead.
\end{abstract}

\begin{highlights}
	\item Ranks APR patches via category-conditioned syntactic frequency patterns.
	\item Improves top-3 correct-patch placement by +27\% over tools’ native order.
	\item Evaluated on 168 Defects4J bugs, 8 tools, 23,032 plausible patches.
\end{highlights}


\end{frontmatter}

\section{Introduction}
\label{sec:introduction}

Despite their significant technical improvements,
automated program repair (APR) techniques remain mostly the
\emph{best effort}:
the patches they produce
come with no absolute guarantees of \emph{correctness}.
In practice, an APR tool
will output several \emph{plausible} (valid)
patches for the faulty program given as input.
Each plausible patch has only been validated
against the test suite (also given as input),
but it may or, more commonly, may not
fully implement the expected, correct behavior for all possible inputs.
The needle-in-a-haystack problem is thus
\emph{ranking} the plausible patches in a way that
those that are correct---if any exist---come before those that are not.
This way, when the developer goes through the list of
patches in ranking order, they will not have to inspect
many plausible fixes until they find one that they can accept as correct.

This paper presents \tool,
a technique that ranks plausible patches, %
trying to put any correct ones higher up in the ranking.
It is common for an APR technique to include patch ranking heuristics
as an integral part of its repair process---for example,
by means of genetic algorithms~\cite{arja}, abstract state enumeration~\cite{pei_automated_2014,pei_jaid_2017},
or similarity as evaluated by a classifier trained on historical fix data~\cite{le_history_2016}.
In contrast, \tool works independent of any patch generation process,
and hence can be used to rank the patches produced by any APR tool.

The key idea of \tool is to rank a patch higher,
the more syntactically similar it is to historic fixes for the same category of bugs.
The historical data comes from mining software repositories
for programmer-written fixes:
for each bug category,
\tool records the syntactic features most frequently used by programmers
to fix those bugs.
For example, it may observe that null-pointer dereferencing errors
are frequently fixed by adding a conditional guard that checks whether
a call's target is \J{null}.
Then, \tool uses this frequency information
to estimate whether a new patch is likely correct.
For example, it will rank patches of null-pointer dereferencing errors
higher (more likely to be correct) if they add a conditional guard to the program,
and lower if they don't.

We designed \tool's training and ranking processes
to be as lightweight as possible:
training %
simply records the relative frequencies of bug category/patch feature combinations;
and ranking directly uses these frequencies to order patches for the same bug.
While the same general idea could be implemented using more sophisticated
machine learning algorithms, we want to demonstrate
that \tool's heuristics are effective even when implemented straightforwardly.
Furthermore, \tool's simplicity ensures that it easily scales
to ranking a large number of patches;
hence, it could be used as a post-processing step of any
APR pipeline with negligible overhead.

We implemented \tool
and used it to rank \numprint{23032} patches
produced by 8 state-of-the-art APR tools for Java
on 168 bugs
in the popular Defects4J curated collection~\cite{just2014defects4j}. While Defects4J includes many more bugs, our evaluation could only consider bugs for which APR tools can produce at least four plausible patches---and hence \emph{ranking} them is a meaningful challenge.
Our experiments indicate that
\tool often successfully
\emph{improves} the ranking of correct patches
over the one produced by the original tool:
it improved the rank of 29\% of the correct patches
from outside to inside the top-3 ranks;
and strictly improved the rank of 43\% of the correct patches.
At the same time, \tool
rarely worsens the ranking of correct fixes
(and, when it does so, it tends to affect fixes that were already ranked very far
from the top positions):
in our experiments,
it worsened the rank of only 2\% of the correct patches
from inside to outside the top-3 ranks.
\tool's effectiveness is largely independent of
the nature of the patches---which APR tool produced them,
and which bugs they repair.
Our comparison with state-of-the-art patch ranking
technique Shibboleth~\cite{Shibboleth2022}
indicates that 
\tool's heuristics remain competitive against %
tools that use more sophisticated machine learning techniques
and include dynamic information (e.g., execution traces).
Overall, these results suggest that \tool
is a widely applicable, scalable, effective technique
to improve the ranking of correct fixes produced by APR systems%
---and thus it can help alleviate the aggravating patch overfitting problem.

\subsection{LLMs for APR: Capabilities and Limitations}
\label{sec:llms-limitations}

As the generative AI juggernaut continues to automate
more and more software engineering tasks,
can \tool's approach remain useful in certain applications?

As noticed in recent surveys~\cite{LbAPR2023,evolving-apr,LLM-apr-survey},
the landscape of APR techniques has broadened greatly.
It is clear that learning-based techniques are increasingly popular%
---especially those based on LLMs (Large Language Models).
At the same time,
different contributions target differ usage (and thus, evaluation)
scenarios,
which complicates comparing their performance and capabilities
across the board.\footnote{
  For example, even when focusing on techniques evaluated on the
  customary Defects4J, ``many systems evaluate on different
  subsets of bugs and mix \passat{k} with accuracy-style metrics''~\cite{LLM-apr-survey}.
}
This highlights a first strength of \tool: since
it is applicable to rank any set of (plausible) patches%
---independent of how the patches were generated in the first place---%
and it is very lightweight in terms of computational resources
(as we discuss in quantitative terms below),
the entry barrier for using \tool is generally quite modest,
which contributes to its flexibility.

On the other hand, if we extrapolate from the pace of progress of LLMs
for automating software engineering tasks,
we may be tempted to conclude that there won't be much room
for traditional techniques such as \tool.
There is clear evidence, however, that there remain
practical scenarios 
where the applicability of LLMs is limited by constraints
of various kinds:
\begin{itemize}
\item \emph{Scalability:}
  as highlighted in a recent survey~\cite{LLM-apr-survey},
  even state-of-the-art APR techniques based on LLMs
  are often only demonstrated on fixing bugs with so-called
  perfect fault localization. This means that the location
  where to apply a patch is given as input to the APR system.
  In contrast, applying APR ``in the wild'' would require
  end-to-end automation---from detecting a bug to devising a fix for it.
  There has been plenty of progress to support
  larger context windows or to summarize a project's codebase
  in a way that it amenable to end-to-end analysis;
  however, the state-of-the-art indicates that there still is a gap
  between research prototypes and industrial-strength tools
  when it comes to APR technology~\cite{LLM-apr-survey}.
  
\item \emph{Cost \& performance:}
  despite significant improvements in open-source and ``small'' models,
  there remains a major performance gap between the LLMs
  that one can run on local hardware (i.e., your laptop) and
  frontier models, which are only accessible in the (commercial) cloud.
  Consider, for example, the latest generation of Qwen models~\cite{qwen3-TR}:
  on the CodeForces coding benchmark~\cite{codeforces},
  the score difference between the best performing \emph{Qwen3-235B-A22B/thinking}
  and the simpler \emph{Qwen3-4B/non-thinking}
  (which can be run on a well-equipped laptop)
  is 64.5 percentage points (98.2\% vs.\ 33.7\%).
  More specifically for APR, the RepairBench benchmark~\cite{repair-bench}
  indicates that the gap between
  the best model (\texttt{gpt-o4-mini}) and the most affordable one (\texttt{mistral-small-2503})\footnote{
    Note that this ``small'' model has 24 billion parameters,
    which means that it still is too big to be run on standard hardware
    with good performance.
  }
  is 30 percentage points (50.3\% vs.\ 20.4\%).\footnote{
    These numbers are from
    the 2025-06-11 snapshot of
    RepairBench's leaderboard at \url{https://repairbench.github.io/}
    (the latest update at the time of writing).
  }
  Thus, state-of-the-art performance requires access to frontier models,
  which incur significant costs for regular usage~\cite{kipf2026waiting}.

  \begin{table}[!htb]
    \centering
    \footnotesize
    \begin{tabular}{clr}
      \toprule
      & \textsc{technique} & \textsc{avg.~\$ per fixed} \\
      \midrule
      \multirow{3}{*}{LLM-based}
       & ChatRepair~\cite{ChatRepair2024} & 0.42 \\
       & RepairAgent~\cite{RepairAgent2025} & 0.14 \\
      & RepairBench~\cite{repair-bench} & 0.20--0.50 \\
      \cmidrule(lr){1-3}
      \multirow{2}{*}{traditional}
        & Jaid~\cite{jaid-TSE} & 0.07 \\
      & Restore~\cite{restore} & 0.04 \\
      \bottomrule
    \end{tabular}
    \caption{A comparison of the costs to fix a bug with different APR technologies.}
    \label{tab:cost-comparison}
  \end{table}

  To better gauge the difference in actual cost between different
  kinds of techniques, \Cref{tab:cost-comparison} summarizes
  the cost per correctly fixed bug reported in the experimental evaluations of different tools.
  ChatRepair~\cite{ChatRepair2024}
  and RepairAgent~\cite{RepairAgent2025} are state-of-the-art LLM-based APR approaches,
  whereas RepairBench~\cite{repair-bench}
  is a leaderboard that compares frontier LLM models applied to APR.
  While each publication may compute the costs in a slightly different way (and LLM costs keep on changing), it is clear that the cost per fixed bug hovers in the \$0.1--\$0.5 range.
  Jaid~\cite{jaid-TSE} and Restore~\cite{restore} are two traditional
  APR techniques we developed in previous work; their cost per fixed bug is an order of magnitude less than the LLM-based tools: \$0.04--\$0.07.
  Finally, let us consider \tool. We call it a ``lightweight'' tool
  because its running costs are negligible compared to the actual costs to produce plausible fixes. As we discuss in \Cref{sec:experiments}, ranking $n$ fixes takes on average $n \times 0.03$ seconds; using the same cloud-computing costs applied
  to Jaid's and Restore's experiments, this translates to
  a cost of \$$10^{-7}$ per ranked fix. Concretely, this means
  that \tool's overhead is simply negligible.
  
\item \emph{Security \& privacy:}
  even when the costs of accessing frontier models are not a problem,
  security and privacy concerns may prevent using LLMs hosted
  by a third-party organization.
  In addition to the security issues inherent in using cloud computing
  services, LLMs present
  specific risks for data privacy~\cite{llm-beyond-privacy,LLM-privacy,ai-privacy-eu}.
  Given that there are limits to the models that can be run on private hardware
  (see previous point), the tension between security and performance
  can complicate the adoption of APR techniques based on LLMs.

\item \emph{Customization:}
  frontier models are commonly showcased on benchmarks using so-called
  high-resource languages: widely popular languages like Python, JavaScript,
  and Java, which feature prominently in the LLMs' vast training data.
  In contrast, researchers have shown that the capabilities of even the
  best models degrade significantly on low-resource languages~\cite{GiagnorioMB25};
  for example, the \passat{1} score~\cite{passatk}
  on code generation tasks decreases, on average, by 38.6 percentage points
  when going from Python (a high-resource language)
  to Racket (a low-resource language, which still has a considerable user base).\footnote{\url{https://www.tiobe.com/tiobe-index/}}
  These data imply that APR techniques based on LLMs
  may be difficult or impractical to customize
  to work reliably on languages that are not as popular as Python or Java.
  This, in turn, highlights other scenarios where traditional APR techniques
  may be preferable because they are language-agnostic and offer more flexibility.
\end{itemize}

In summary, while researchers and practitioners are busy at work
trying to overcome some of the aforementioned limitations,
traditional APR techniques still have something to offer
in certain application domains.

\subsection{Positioning}
\label{sec:positioning}

\tool is technique-agnostic, in that it can
be applied to rank the plausible patches produced by any
APR technique, regardless of how the technique works.
The experimental evaluation that we present in \Cref{sec:experiments}
mainly applies \tool to patches generated by traditional APR tools
that are not based on LLMs.
This is simply because most recent LLM-based APR techniques 
follow an iterative generation process,
which rarely produces more than one plausible fix per bug;
under these conditions, ranking is immaterial.
This is not an intrinsic limitation of \tool,
as it remains applicable to rank any kind of patches regardless of how they were generated;
in fact, \Cref{sec:rq1} discusses a small experiment where
\tool ranked the patches produced by a zero-shot prompted LLM.

It is clear that
LLM-based techniques are lionized in recent APR research,
since they allow researchers to develop novel approaches
that leverage the capabilities of frontier generative AI models.
At the same time,
traditional APR methods---based on structured search and template-based techniques---still work well for certain bugs
that can be fixed with small, local syntactic changes (e.g.,
adding null checks or adjusting a conditional)~\cite{TGPR2025}.
Therefore, such traditional methods remain useful as baselines and in highly constrained systems~\cite{evolving-apr};
in other words, they work well as specialized tools.
As for LLM-based APR methods, while they are often general purpose,
it has been recently shown that the capabilities of different agentic systems for program repair
are often complementary, as ``no single agent consistently outperforms others''~\cite{agentic-comparison}.
A similar complementarity has been observed in various test-driven LLM-based APR approaches~\cite{PReMM2025}.
Finally, \Cref{sec:llms-limitations} presented several scenarios where
techniques based on state-of-the-art LLMs are impractical because they are too expensive or
fail to satisfy safety and privacy constraints.

These considerations suggest that
the landscape of APR techniques remains varied
and populated by complementary solutions that are applicable in different conditions.
As a result, \tool's approach remains useful too:
\begin{enumerate*}
\item it provides a practical way to improve the accuracy of APR systems with a lightweight, inexpensive technique
  that is applicable across the board; and
\item it supports a simple way of integrating multiple APR approaches to leverage their complementary fixing capabilities.
\end{enumerate*}

\subsection{Contributions}

In summary, this paper makes the following contributions:
\begin{enumerate}
\item \tool: a novel technique for ranking plausible APR patches, and its
  implementation. %
\item An experimental evaluation of \tool
  ranking plausible patches produced by state-of-the-art APR tools for Java,
  suggesting that \tool is often effective, and compares
  favorably to other APR patch ranking techniques.
\item The implementation of \tool and all details of the experimental evaluation,
  available as a replication package (see~\cref{sec:data-availability}). 
\end{enumerate}

\subsection{Discussion of Implications}
\label{sec:implications}

Let us wrap up the introduction with a brief discussion of
the implications of the research results obtained described in the rest of the paper.

The main selling point of \tool for \emph{practitioners}
is that it is a very affordable technique,
which runs on basic hardware and is applicable to rank plausible
patches regardless of the technique used to generate them.
Whether \tool's effectiveness---how much it successfully
improves the ranking of correct fixes---is sufficient in practice
depends, of course, on the specific requirements of the concrete application
scenarios.
It remains that APR is not a solved problem~\cite{LLM-apr-survey},
and the state-of-the-art in this field often advances
in dribs and drabs;
hence, \tool's capabilities are a valuable, if incremental, result.

For \emph{researchers},
\tool's work indicates that exploring APR techniques
that are not based (primarily or exclusively) on LLMs
remains an interesting research problem,
which may lead to complementing the state-of-the-art,
and cater to the needs of users with different requirements
(see the constraints discussed in \Cref{sec:llms-limitations}).
Another suggestion is that
allowing an APR tool to explicitly output multiple plausible
patches for the same bug may suggest novel opportunities
of combining repair with ranking techniques,
which may open up new paths to research progress in automated program repair.

\section{How \tool Works}
\label{sec:approach}

\paragraph{Background: How APR Works}
Automated program repair (APR) 
encompasses a wide variety of techniques aimed at producing,
completely automatically, source-code fixes for programming errors.
The input of an APR tool usually includes
the source code of a buggy program,
as well as a test suite $T$ that is used as a (partial) specification
of the expected correct program behavior;
thus, $T$ should include at least one failing test,
which exposes the bug to be repaired.

The output of an APR tool consists of
source-code patches that ``repair'' the buggy program
and pass all tests in $T$.
Such patches are called \emph{plausible} because they have been
validated (at least) against the test suite $T$.
However, they may or may not be actually \emph{correct}:
since a test suite only checks a finite number of possible program behaviors,
it is possible that a plausible patch is actually incorrect
because the patched program still behaves incorrectly
in scenarios not checked by any of the available tests in $T$.

Since complete and formal specifications of software are rarely available,
the ultimate judge of correctness is the developer,
who can use their intuition and knowledge of the codebase
to determine whether a certain program behavior is acceptable or wrong.
Correspondingly, APR tools are usually evaluated on historic bugs,
using the programmer-written fix for a certain bug as
the yardstick to evaluate correctness.

Another consequence of the under-specification provided by tests
is that an APR tool may produce different plausible patches
for the same bug.
This motivates the usefulness of \emph{ranking} plausible patches
according to their likelihood of being correct.

In the following, we use the term \emph{fix} only to denote
a programmer-written repair that is considered correct.
In contrast, the term \emph{patch} denotes
any program modification that is produced by an APR
system; thus, a patch may or may not be correct.

\subsection{An Overview of \tool}
In a nutshell,
\tool is a technique to rank plausible patches produced
by automated program repair tools
based on how they are similar to
programmer-written fixes to historic bugs with
comparable characteristics.
Thus, the \tool approach consists of two phases:
a training phase that learns from historic bugs and fixes,
and a ranking phase that uses the learned frequencies
as heuristics to rank newly produced plausible patches.

\paragraph{Training}
\cref{fig:learning} overviews \tool's training phase,
whereas \cref{alg:training} illustrates it with pseudo-code.
In this phase,
\tool mines software repositories for historical data about
bugs and their corresponding programmer-written fixes,
in order to learn which features of fixes
are more commonly associated with certain bug categories.
\tool classifies each bug $b$ in a \emph{category} $\bugc[b]$
based on how it is described by the developers that identified and fixed it
in the related commit messages;
examples of bug categories include ``overflow'' and ``null-pointer'' bugs.
\tool also classifies each programmer-written fix $f$ in a \emph{kind} $\patchk[f]$
based on its syntactic features, that is how it modifies the buggy source code.
Examples of features combined to define fix kinds include
``adding a conditional'', ``changing the return type'', and ``initializing a local variable''.
We curate
bug categories and fix kinds so that they are mutually exclusive: %
any bug belongs to exactly one category,
and any fix belongs to exactly one kind
(a combination of several syntactic features).
\tool summarizes this information by reporting
\emph{frequencies} $\freq[c][k]$:
the fraction of all bugs of category $c$ in the training set
whose programmer-written fix is of kind~$k$.

\begin{figure*}
  \centering
  \begin{adjustbox}{width=\linewidth}
  \begin{tikzpicture}

    \node[label={[align=center]south:\textbf{frequencies} \freq[c][k]}]
    (distro) {\freqDistro{20mm}};

    \node[below=of distro,
    label={[align=center]south:\textbf{plausible} \textbf{patches}}]
    (plausible) {$p_1\text{\textcolor{gray}{\faWrench}}
      \ \:
      p_2\text{\textcolor{gray}{\faWrench}}\ \ldots
      \ \:p_n\text{\textcolor{gray}{\faWrench}}$};

    \node [fit=(plausible)(distro),draw=none] (inputs) {};
    
    \coordinate (mid-left) at ($(distro)!0.75!(plausible)$);
    \coordinate (join-left) at ($(4mm,0)+(mid-left -| inputs.east)$);
    \coordinate[above=of join-left] (join-left-2);

    \node[right=21mm of join-left]
    (patches) {$k_1\text{\textcolor{hcol}{\faWrench}}
      \ \:
      k_2\text{\textcolor{gcol}{\faWrench}}\ \ldots
      \ \:k_n\text{\textcolor{gcol}{\faWrench}}$};

    \node[above=5mm of patches]
    (bug) {$\bugc*$\:\textcolor{acol}{\faBug}};
    
    \draw[->,thick] (distro.east) -| (join-left-2) -- node[above,align=center]
    {\textcolor{rankcol}{\textbf{bug class}}\\\textcolor{rankcol}{\textbf{estimation}}}
    (join-left-2 -| bug.west);
    \draw[thick] (plausible.east) -| (join-left-2);

    \draw[->,thick] (join-left) -- node[below,align=center]
    {\textcolor{rankcol}{\textbf{patch}}\\
      \textcolor{rankcol}{\textbf{classification}}}
    (join-left -| patches.west);

    \coordinate (mid-right) at ($(2mm,0)+(patches.east)!0.5!(bug.east -| patches.east)$);
    \node[right=20mm of mid-right,align=center,
    label={[align=center]south:\textbf{ranked} \textbf{patches}}]
    (ranked)
    {$p_{r_1}\text{\faWrench}$
      \\$p_{r_2}\text{\faWrench}$
      \\$\vdots$
      \\$p_{r_n}\text{\faWrench}$};

    \draw[->,thick] (bug.east) -| (mid-right) --
    node[align=center]
    {\textcolor{rankcol}{\textbf{frequency}}\\\textcolor{rankcol}{\textbf{ranking}}}
    (ranked.west);
    \draw[thick] (patches.east) -| (mid-right);
  \end{tikzpicture}
\end{adjustbox}
  \caption{An overview of \tool's \emph{ranking} phase:
    \tool determines the \emph{kind} of each plausible patch
    (generated by a program repair tool), using the same heuristics used in the
    training phase;
    based on the frequency statistics collected in the training phase,
    it also \emph{estimates} the \emph{category} $\overline{c}$
    of the bug these patches
    are fixing;
    finally, it ranks each plausible patch of kind $k$
    as per the frequency distribution $\freq[\overline{c}][k]$.
    }
  \label{fig:ranking}
\end{figure*}
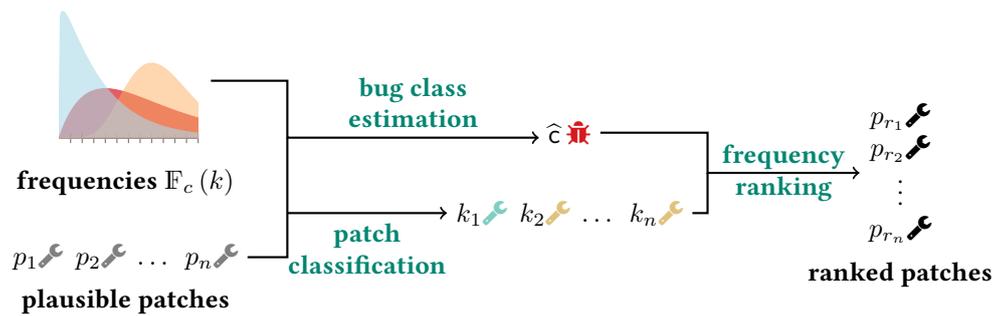

\paragraph{Ranking}
\cref{fig:ranking} overviews \tool's ranking phase.
In this phase,
\tool uses the frequency information collected in the training phase
as heuristics to rank a set $P$ of plausible patches produced
by an automated program repair tool for some bug $b$.
First, \tool classifies each patch $p \in P$ in a \emph{kind} $\patchk[p]$;
the classification of plausible patches is exactly as the
classification of programmer-written fixes,
since it is based entirely on the patch's syntactic features.
Then, %
\tool \emph{estimates} $b$'s bug category
as the category $\bugc*[b]$
for which patches of the same kinds as those in $P$ are more frequent.
Since the classification of bugs into categories
uses information that is generally only available \emph{after} a bug
has been fixed,
\cref{eq:category-estimate} is an educated guess;
however, our experiments will show that this heuristics is often reliable.
Finally, %
\tool
ranks each patch $p \in P$
by decreasing values of $\freq[\bugc*[b]][\patchk[p]]$.
Intuitively,
the more frequent a fix of that kind was in the historical data
for the estimated bug category,
the higher it is ranked over the other patches.

The rest of this section describes the main steps of \tool's approach
in detail, and discusses other aspects of its current implementation.

\begin{figure*}
  \centering
  \begin{adjustbox}{width=\linewidth}
  \begin{tikzpicture}
    \begin{scope}[node distance=3mm and 5mm]
      \node (bug1) {$b_1$\faBug\ \:$f_1$\faWrench\ \:$m_1$\faStickyNoteO};
      \node[below=of bug1]  (bug2) {$b_2$\faBug\ \:$f_2$\faWrench\ \:$m_2$\faStickyNoteO};
      \node[below=of bug2]  (bug-dots) {$\cdots$};
      \node[below=of bug-dots]  (bugn) {$b_n$\faBug\ \:$f_n$\faWrench\ \:$m_n$\faStickyNoteO};
    \end{scope}
    \node [fit=(bug1)(bug2)(bug-dots)(bugn),draw=none,
    label={south:\textbf{bug/fix history}}] (history) {};

    \node[left=15mm of history,
    label={[align=center]south:\textbf{project}\\\textbf{repositories}}]
    (repos) {\Huge\faGithub};

    \draw[<-,thick] (history.west) -- node[above] {mining} (history.west -| repos.east);

    \begin{scope}[node distance=3mm and 5mm]
      \node[right=30mm of bug1] (class1)
      {$c_3$\textcolor{bcol}{\faBug}\ \:$k_4$\textcolor{gcol}{\faWrench}};
      \node[below=of class1] (class2)
      {$c_2$\textcolor{acol}{\faBug}\ \:$k_1$\textcolor{hcol}{\faWrench}};
      \node[below=of class2]  (class-dots) {$\cdots$};
      \node[below=of class-dots] (classn)
      {$c_1$\textcolor{dcol}{\faBug}\ \:$k_4$\textcolor{gcol}{\faWrench}};
    \end{scope}

    \node [fit=(class1)(class2)(class-dots)(classn),draw=none,
    label={[align=center]south:\textbf{bug categories}\\\textbf{fix kinds}}] (classification) {};

    \coordinate (h-c-m) at
    ($(history.east)!0.15!(history.east -| classification.west)$);

    \draw[->,thick] (history.east) -- (h-c-m)
    |- node[above,near end,align=center] {\textcolor{rankcol}{\textbf{bug}}\\\textcolor{rankcol}{\textbf{classification}}}
    ($(classification.west)+(0,5mm)$);

    \draw[->,thick] (history.east) -- (h-c-m)
         |- node[below,near end,align=center] {\textcolor{rankcol}{\textbf{fix}}\\\textcolor{rankcol}{\textbf{classification}}} ($(classification.west)+(0,-5mm)$);
         
    \coordinate (c-freq) at (history.east -| classification.east);

    \node[right=14mm of c-freq,
    label={[align=center]south:\textbf{frequencies} \freq[c][k]}] (distro) {\freqDistro{30mm}};
    
    \draw[->,thick] (classification.east) -- node[above] {statistics} (classification.east -| distro.west);
  \end{tikzpicture}
\end{adjustbox}
  \caption{An overview of \tool's \emph{training} phase:
    by mining historic data in software repositories,
    \tool builds distributions \freq[c][k]
    that summarize how frequently a certain patch kind $k$ was used by
    developers to fix a certain bug category $c$.}
  \label{fig:learning}
\end{figure*}
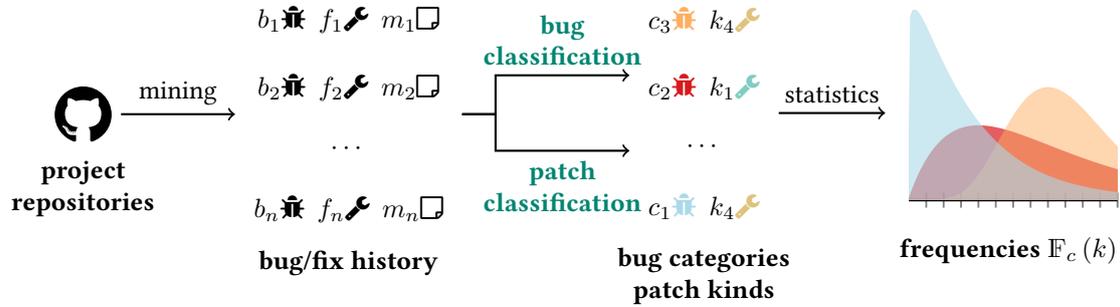

\subsection{Training Data}
\label{sec:training-data}

\tool's training phase
mines historical data about bugs and their programmer-written fixes
in GitHub repositories.
For each pair $c^-, c^+$ of consecutive commits in a project repository's
main branch, \tool considers the triple $\langle b, f, m\rangle$ where:
\begin{enumerate*}
\item $b$ is the source code in commit $c^-$;
\item $f$ is the ``diff'' of $c^+$ over $c^-$;
\item $m$ is $c^+$'s commit message.
\end{enumerate*}

\tool determines whether to further consider a triple $\langle b, f, m\rangle$
according to some heuristics.
First, it discards the triple if $f$ involves (adds, removes, or modifies)
more than five lines of code:
since generating complex multi-line repairs
is generally outside the current capabilities of APR,
including complex patches would make our training data
less representative of the bugs and fixes that we want to be able to rank.
Second, \tool only retains the triple if its commit message $m$
includes keywords that suggest that it is a bug-fixing commit
(as opposed to a commit that adds or changes functionality,
or that affects only tests or documentation).

\subsection{Bug Classification}
\label{sec:classify-bug}

Each bug $b = \langle b, f, m \rangle$ in the training data
is labeled with a \emph{category} $\bugc[b]$.
To this end, we introduced a classification
based on a small number of well-defined bug categories:
too many fuzzy bug categories would dilute the ``signal'' in the training data,
and would be much harder to heuristically match in the ranking phase
(see~\cref{sec:estimate-bug}).
Based on some exploratory analysis%
---as well as on common bug categories used in curated collections
such as Defects4J~\cite{defects4j,just2014defects4j}---%
we decided to focus on three widely applicable, well-known, and distinct 
bug categories: \emph{overflow} bugs, \emph{null-pointer} dereference bugs, and \emph{logic} bugs (i.e., bugs that do not crash the program but involve the violation of a logic condition, such as an assertion failure).
Overall, these three categories cover a diverse range of different bugs
that are within the purview of APR's state-of-the-art;
this helps make \tool applicable in combination with several
different APR tools---which we'll demonstrate in the experiments.

For the whole ranking process to work,
the training data has to be categorized accurately (i.e., no misclassifications);
otherwise, the training phase would end up with spurious associations
between bugs and patches.
To this end, we categorized each bug in a semi-automated way:
first, we used a keyword-based search to assign a \emph{tentative} category
to each bug automatically; then,
we manually reviewed and validated the keyword-based classification.
The first, automated step speeds up the second, manual step
and makes the classification effort reasonable.
Furthermore, note that this classification need only be done once
to prepare the training data for \tool;
using \tool to rank the patches produced by an APR tool
remains a fully automated process.
The rest of the section details the classification process.

\paragraph{Keyword-based search}
For each category, we defined a number of keywords that may indicate
a bug of that category;
For example ``buffer'' and ``overflow'' may indicate overflow bugs,
whereas ``NPE'', ``null'', and ``pointer'' may indicate null-dereference bugs;
\cref{code:regex} details the regexes we used in our experiments.
If the commit message $m$ of a bug $b$ matches
the regex associated with category $c$,
then we tentatively assign category $c$ to bug $b$.
If more than one regex matches, the manual validation step
will determine if any of the corresponding categories is correct,
or if it is preferable to omit the bug from the training set
to avoid ambiguity.
In our experiments, we found 167 cases of bug commit messages
matching more than one regex, which we excluded from the training set.
  
\begin{lstlisting}[caption=Regex patterns used to identify bug \emph{categories},language=Java,label={code:regex}, autogobble=true,numbers=none,basicstyle=\ttfamily\footnotesize, float]
LOGIC_BUGS_PATTERN = ".*\b(logic|logical)\b.*"
		+ ".*\bfix|bug|issue|wrong|error|fault|assert|correct|condition|unexpected|\b(incorrect|wrong)\s+(function|output|result)\b.*|"
NULL_BUGS_PATTERN = "(?i)((null\s*pointer|NPE|NullPointerException))"
OVERFLOW_BUGS_PATTERN = "(?i)(.*\b(buffer|array)\b.*\b(overflow)\b.*|"
    + ".*\bout\sof\s(bounds|range|limit)\s.*|"
    + ".*\b(heap|stack|integer)\s(overflow)\b.*|"
    + "(Array|String|Index)OutOfBoundsException.*|"
    + "\b(BufferOverflowException\b.*)\n)"
\end{lstlisting}

\paragraph{Manual validation}
The keyword-based search applies simple heuristics
that may misclassify a bug;
we only used as a starting point for a systematic manual validation
of the classification.
To this end,
two authors independently inspected each bug $b$,
together with its patch $f$, commit message $m$
and any other information related to each data point with a keyword-based match,
in order to ascertain whether the automatic classification was correct.
If both authors accept the automatic classification,
the data point $\langle b, f, m \rangle$ is retained and the bug $b$
is classified of category $\bugc[b]$;
otherwise, the data point is removed from the training data.
Henceforth, ``training data'' refers to the culled training data
where every bug has validated category.

\subsection{Patch Classification}
\label{sec:classify-patch}

\tool assigns a kind $\patchk[f]$ to the fix $f$
in each triple $\langle b, f, m \rangle$ in the training data.
Informally, a patch kind represents
one class of syntactic modifications that are present in $f$.
\begin{table}
  \centering
  
  \begin{adjustbox}{width=\linewidth}
  \begin{tabular}{l|lll}
    \toprule
    \multicolumn{1}{c}{\textsc{feature}}
    & \multicolumn{1}{c}{\textsc{add}}
    & \multicolumn{1}{c}{\textsc{remove}}
    & \multicolumn{1}{c}{\textsc{modify}}
    \\
    \midrule
    assignment & assignment & assignment & target, expression
    \\
    conditional & conditional, else & conditional, else & condition (strengthen, weaken, other)
    \\
    loop & loop, \J{break}, \J{continue} & loop, \J{break}, \J{continue} & condition (strengthen, weaken, other)
    \\
    method call & call & call & callee, call arguments
    \\
    \J{return} & \J{return} & \J{return} & returned value
    \\
    \J{try/catch} & \J{catch}, \J{finally} & \J{catch}, \J{finally} & exception type
    \\
    \J{synchronized} block & block & block & lock object, block
    \\
    field declaration &  field & field & signature, initialization
    \\
    method declaration & method & method & signature
    \\
    \J{assert} & assertion & assertion & predicate (strengthen, weaken, other)
    \\
    \bottomrule
  \end{tabular}
\end{adjustbox}
  \caption{The main feature/modification combinations that \tool uses to classify patches. A \textsc{feature} is a syntactic feature of Java code; a modification \textsc{add}s a certain feature, \textsc{remove}s it, or \textsc{modifies} some of its components.} 
  \label{tab:patch-kinds}
\end{table}
\raggedbottom

To come up with a general selection of patch kinds,
we first considered common \emph{syntactic features}
that may be involved in Java patches.
\cref{tab:patch-kinds}'s leftmost column lists the main ones we considered for this work,
including assignments, conditionals (\J{if}), loops (\J{for}, \J{while}),
as well as class-level declarations.
For each syntactic feature,
we considered its main components,
and how a patch that adds, removes, or modifies code may affect them.
\cref{tab:patch-kinds}'s other columns
list several such \emph{combinations},
such as ``\textsc{add} assignment'' (the patch introduces a new assignment),
``\textsc{remove} \J{catch}'' (the patch removes a \J{catch} block from an existing \J{try}/\J{catch}),
and ``\textsc{modify} method signature'' (the patch changes the types or the visibility of an existing method).
This process identified 67 such combinations of feature/modification;
for brevity, \cref{tab:patch-kinds} lists only a representative subset.
The list we compiled is consistent with empirical studies of
the most common bug-fix patterns that are found in open-source Java software~\cite{fix-patterns,fix-patterns-extended,change-patterns,bugfixpatterns}.\footnote{
  For example, the five most common categories of bug-fix patterns in Pan et al.'s manual analysis~\cite{bugfixpatterns}
  correspond to
  ``modify condition in conditional'' (IF-CC in~\cite{bugfixpatterns}),
  ``modify call arguments in method call'' (MC-DAP and MC-DNP),
  ``modify expression in assignment'' (AS-CE),
  and ``add conditional'' (IF-APC)
  in \autoref{tab:patch-kinds}'s classification.
}
\cref{code:fix-conditional} shows
an example of patch from project Apache Tomcat\footnote{\url{https://github.com/apache/tomcat/commit/e886ee20183b5f5d2a902843b6d324b58d991ea8}}
that targets a \emph{conditional} by \emph{modifying} (precisely, \emph{strengthening})
its \emph{condition}.

A patch \emph{kind} is any element of the power set $K = 2^{M}$,
where $M$ is the aforementioned set of all feature/modification combinations.
Since a given patch may involve more than one modification,
this definition of $K$ ensures that each fix $f$ belongs to
a \emph{unique} kind $\patchk[f]$.
Combining features into kinds is key to having a sufficiently fine-grained,
and thus discriminating, classification of patches.

\begin{lstlisting}[caption={An example of programmer-written fix classified as ``\textsl{conditional: modify condition (strengthen)}''. The diff is shown in \textcolor{blue}{blue} and \underline{underlined}.}, label={code:fix-conditional}, basicstyle=\footnotesize,float=tb]
 if ((*\underline{\color{blue}buffered.getBuffer() != null \&\&}*)(* \footnotesize buffered.getBuffer().length > 65536) *)
\end{lstlisting}

\subsection{Historic Frequencies}
\label{sec:history-frequency}

As described in the previous sections,
\tool assigns a bug category \bugc[b]
and a patch kind \patchk[f]
to the bug $b$ and fix $f$
in each triple $\langle b, f, m \rangle$
in the training data $T$.
Based on this classification,
\tool summarizes the associations between
bugs and fixes by means of a
historic \emph{frequency} $\freq[c] \colon K \to [0, 1]$
for each bug category $c \in C$.
Given a unique fix kind $k \in K$ and bug category $c \in C$,
\freq[c][k]
is the fraction of all triples in the training data
where a bug of category $c$ was fixed with a patch of kind $k$:
\begin{equation}
  \label{eq:frequencies}
  \freq[c][k]
  \ =\
  \frac
  { |\{ \langle b, f, m \rangle \in T \mid \bugc[b] = c, \patchk[f] = k \}| }
  { |\{ \langle b, f, m \rangle \in T \mid \bugc[b] = c \}| }
\end{equation}
You can think of \freq[c][k]
as a normalized frequency distribution that captures how frequently
a fix of kind $k \in K$ was used historically to repair
a bug of category $c \in C$.
These frequencies are the overall output of \tool's training phase.

\begin{algorithm}[tb]
	\small
\caption{\tool's training phase.}
\label{alg:training}
\KwIn{An ordered sequence of consecutive commits $c_1, c_2, \ldots, c_m$}
\KwOut{A collection of frequencies $\freq \colon C, K \to [0, 1]$}

\tcp{\# commit pairs of bug category $c \in C$ and fix kind $k \in K$}
$N \colon C, K \to \mathds{N} \leftarrow 0$ \qquad \tcp{initialized to all zeros}
\ForEach{$x \in \{1, \ldots, m - 1\}$}{   \label{algo1:line2}
  $c^-, c^+ \leftarrow c_x, c_{x+1}$  \qquad \tcp{pair of consecutive commits}
  $b, f, m \leftarrow \mathsf{code}(c^-), \mathsf{diff}(c^-, c^+), \mathsf{message}(c^+)$ \\
  \If{$\mathsf{size}(f) \leq 5 \land \mathsf{is\_bug\_fix}(m)$}{ \label{algo1:line5}
      $\bugc \leftarrow \mathsf{bug\_category}(m)$ \qquad \tcp{bug classification}
      \tcp{human validation of bug category}
      \If{$\mathsf{valid}(b, f, m, \bugc)$}{
          $\patchk \leftarrow \mathsf{patch\_kind}(f)$ \qquad \tcp{patch classification}
          $N(\bugc, \patchk) \leftarrow N(\bugc, \patchk) + 1$ $\ \;$\tcp{increment count}
        }
	}
 }
\tcp{Compute frequencies for each $c \in C$ and $k \in K$}
\ForEach{$c \in C$}{
  \ForEach{$k \in K$}{
    $\freq(c, k) \leftarrow N(c, k) / \sum_{k \in K} N(c, k)$
  }
}
\Return \freq
\end{algorithm}

\subsection{Bug Category Estimation}
\label{sec:estimate-bug}

In the ranking phase,
\tool inputs the frequencies $\freq[c]$
produced by the training phase,
and a collection $P$ of patches
produced by an APR system for a certain bug $b$.
\tool assigns a kind $\patchk[p]$
to each patch $p \in P$
according to the same algorithm used in the training phase
to classify fixes (\cref{sec:classify-patch}).
However, it cannot classify the \emph{bug} under repair
in the same way it classified bugs in the training phase:
bug classification (\cref{sec:classify-bug})
relies on the commit message
that accompanies a programmer-written fix---as well as, possibly,
on other information.
In contrast, the bug category should be \emph{estimated automatically}
in the ranking phase, which works on automatically generated patches
whose correctness and intent are unknown in general.

\tool estimates $b$'s bug category $\bugc*[b]$ as the category that
was most frequently associated with the kinds of patches $P$
in the training data:
\begin{equation}
  \bugc*[b] \ =\ \argmax_{c \in C} \sum_{p \in P} \freq[c][\patchk[p]]
\label{eq:category-estimate}
\end{equation}
Since we only have to distinguish between a small number of bug categories,
this simple heuristics is generally robust and
leads to effective results---as demonstrated in \cref{sec:experiments}.
\paragraph{Tie-breaking}
If two or more categories obtain the same score in Eq.~(2), we pick the
category with the larger number of training instances $|T_c|$; if still tied, we
choose the one with larger $\max_{p\in P} F_c(k(p))$; remaining ties are
resolved by a fixed (lexicographic) category order to keep the procedure
deterministic.

\subsection{Patch Ranking}
\label{sec:rank}

In the final step of the ranking phase,
\tool ranks each patch $p \in P$ among those produced by
an APR system for a bug $b$ according to the value
$\freq[\bugc*[b]][\patchk[p]]$: the higher this value is for $p$
relative to the other patches in $P \setminus \{p\}$,
the higher $p$ is ranked by \tool.

An important detail is how to handle \emph{ties} in the ranking.
When several patches %
all have the same value of $\freq[\bugc*[b]][\patchk[p]]$,
then \tool sorts them
in the same relative order as the APR system's original ranking.
For example, if four patches have score $p_3=0.7, p_1 = p_4 = 0.5, p_2=0.3$,
and the original APR system ranks them in the higher-to-lower order $p_4, p_3, p_2, p_1$,
\tool ranks them as $p_3, p_4, p_1, p_2$.
Note that ties are not very frequent, thanks to the
fine-grained classification into patch kinds.

\section{Experimental Evaluation}
\label{sec:experiments}

\tool's experimental evaluation addresses the following
research questions:
\begin{description}[leftmargin=10mm]
\item[RQ1] How \emph{effective} is \tool's ranking of patches? \\
  This RQ investigates how frequently \tool ranks \emph{correct} patches
  higher than plausible but incorrect ones.
  
\item[RQ2]  How does \tool \emph{compare} to
  other state-of-the-art patch ranking techniques? \\
  This RQ compares \tool to other approaches
  to patch ranking, focusing on effectiveness and complementarity.

\item[RQ3] How \emph{robust} is \tool's classification of bugs and patches? \\
  This RQ investigates whether  \tool's approach is \emph{robust},
  that is how the amount and variety of
  historical data that is available for training
  affects \tool's ranking performance.

\item[RQ4] Can \tool's manual training steps be fully \emph{automated}?\\
  This RQ investigates whether it is feasible to use LLMs
  (Large Language Models) to automate the bug identification and classification
  steps of \tool's training.
\end{description}
The rest of this section describes the experimental setup 
and the answers to these research questions.

\begin{table}
  \centering
  \begin{subtable}{\textwidth}
    \centering
    \small
    \begin{tabular}{l rrrr}
    \toprule
    & \textsc{size} \textsl{(kLOC)}  & \textsc{commits} & \textsc{stars} & \textsc{fixes} \\
    \midrule
    \textbf{mean} & \numprint{2303} & \numprint{20567} &  \numprint{9725} & \numprint{81} \\
    \textbf{max} & \numprint{22500} & \numprint{103949} & \numprint{68900} & 578 \\
      \textbf{min} & \numprint{34} &  \numprint{2076} & \numprint{46} & \numprint{2} \\
      \cmidrule(lr){2-5}
    \textbf{total} & \numprint{186533} & \numprint{1665958} & \numprint{787737} & \numprint{6583} \\
    \bottomrule
  \end{tabular}
  \caption{Characteristics of the 81 Java open-source projects used as training data.
    The table reports the average (mean), maximum, minimum, and total value
    of the projects' \textsc{size} (in thousands of lines of code),
    number of \textsc{commits} (on the main branch), 
    number of GitHub \textsc{stars},
    and number of bug \textsc{fixes} that we included in our training data.
  }
  \label{tab:subjects-stats}
\end{subtable}
 \begin{subtable}{\textwidth}
 	\centering
 	\small
\begin{tabular}{llrrrr}
	\toprule
	\multicolumn{1}{c}{\textsc{project}} &
	\multicolumn{1}{c}{\textsc{version}} &
	\multicolumn{1}{c}{\textsc{mloc}} &
	\multicolumn{1}{c}{\textsc{commits}} &
	\multicolumn{1}{c}{\textsc{kstars}} &
	\multicolumn{1}{c}{\textsc{\% Java}} \\
	\midrule
	spring-boot   & v3.1.2  & 0.80 & \numprint{44563} & 69 & 98\% \\
	elasticsearch & 8.9.1   & 4.60 & \numprint{71418} & 65 & 100\% \\
	spring        & v6.0.11 & 1.50 & \numprint{27652} & 53 & 98\% \\
	guava         & 32.1.2  & 1.00 & \numprint{6155}  & 48 & 100\% \\
	RxJava        & 3.1.6   & 0.48 & \numprint{6042}  & 47 & 100\% \\
	retrofit      & 2.9.0   & 0.04 & \numprint{2076}  & 42 & 96\% \\
	dubbo         & 3.2.5   & 0.45 & \numprint{7008}  & 39 & 99\% \\
	dbeaver       & 23.1.4  & 0.89 & \numprint{24589} & 33 & 100\% \\
	afa           & 3.5.1   & 1.20 & \numprint{11544} & 26 & 77\% \\
	flink         & 1.17.1  & 3.70 & \numprint{33859} & 22 & 86\% \\
	\bottomrule
\end{tabular}

      \caption{The ten largest GitHub projects used by \tool's training phase.The table reports the analyzed project \textsc{version}, the project size in millions of lines of code (\textsc{mloc}), the number of \textsc{commits}, the thousands of GitHub stars (\textsc{kstars}), percentage of source code that is Java (\textsc{\%~Java}), and the number of bug \textsc{fixes} that we included in our training data.}
      \label{tab:top10-subjects}
    \end{subtable}\\[-2mm]
    \label{tab:subjects}
    \caption{Subjects used by \tool's training phase.}
\end{table}

\paragraph{Implementation and performance}
We implemented \tool in Java, comprising 5.3k LOC.
\tool uses GumTree~\cite{gumtree} to analyze patch features, JavaParser~\cite{javaparser} to parse Java code,
and JGit\footnote{\url{https://github.com/eclipse-jgit/jgit}}
to mine GitHub repositories.

Since it is a lightweight, scalable tool,
we did not need to analyze in detail the running-time performance of \tool.
In particular,
\tool's running time %
is roughly proportional to the number of
plausible patches it has to rank:
on average, \tool takes 30 milliseconds per patch
(median: 28 ms, max: 41 ms, standard deviation: 7.5 ms).
As we mentioned in \Cref{sec:introduction},
\tool's scalability and modest computational cost
are key in making it a complementary tool
to approaches based on generative AI,
which are usually heavyweight and only accessible
on specialized hardware/cloud platforms.

\subsection{Training Phase}
\label{sec:training}

\paragraph{Projects}
We extracted the training data
by mining the commit histories of
81 popular open-source projects hosted by GitHub.
We randomly selected these projects among those with
at least 40 stars, 30 thousand lines of code,
and 2 thousand commits.
\Cref{tab:subjects-stats} summarizes the main characteristics
of these 81 projects; \Cref{tab:top10-subjects} lists
the ten largest projects in our selection.

\paragraph{History mining}
\label{sec:training-data}

We mined the commit histories of the selected 81 projects,
in order to extract suitable training data to be used as described in \Cref{sec:training-data}.
We started from all \numprint{1665958} commits in our projects.
We discarded all commits that add, remove, or modify more
than five lines of code; \numprint{224965} commits satisfy this criterion. 
Three considerations guide the decision to restrict patches to five lines of code. First, empirical studies show that real-world bug fixes are typically small: the median patch size in DefectsJS is four lines~\cite{8330203}, and large-scale analyses of production Java projects report that corrective patches rarely exceed 20 lines~\cite{8941305, jrepairanalysis}. Second, existing APR tools, including recent LLM-guided tools, generally produce patches of only a few lines. Third, our patch-kind classifier infers syntactic intent (e.g., null-check addition, condition strengthening). For larger multi-line patches that introduce complex logic or method bodies, such inference becomes unreliable without deeper program context. For these reasons, we focus on small patches, which are both representative of real bug fixes and compatible with the abstraction capabilities of our classifier.

Next, we discarded all \numprint{217181} commits
whose commit message does \emph{not} match any of \Cref{code:regex}'s regular expressions.
As described in \Cref{sec:classify-bug},
we used the regexes as \emph{necessary} conditions,
followed by a
manual inspection that
all regular expression matches are indeed bug-fixing commits;
precisely,
two authors carried out the manual inspection independently,
and we selected a commit,
out of the remaining \numprint{7784} ($= 224965 - 217181$),
only if both agree that it is a genuine
bug-fixing commit, resolving a bug of the category given by the regex
(i.e., overflow, null pointer, or logical).
This conservative process,
which took around two full days of work,
gives us a good confidence that
the selected \numprint{6583} commits do indeed represent bug fixes
of a certain category (\numprint{1093} logical, \numprint{5161} null pointer, \numprint{329} overflow) in a project's history.

\begin{table}[!tb]
	\caption{Ten frequently occurring patch kinds in our training data. For each patch kind, the table reports the percentage of bugs of each category that were fixed by a patch of that kind.}
	\label{tab:category}
	\resizebox{\columnwidth}{!}{%
		\begin{tabular}{lcrr}
        \toprule
        \multicolumn{1}{c}{\textsc{patch kind}}
        & \multicolumn{3}{c}{\textsc{\% bug of category}}         \\
			& \multicolumn{1}{c}{\textsl{logical}}   & \multicolumn{1}{c}{\textsl{null pointer}} & \multicolumn{1}{c}{\textsl{overflow}}  \\ \midrule
			\textit{add} conditional                        & 5  & 13    & 6 \\ 
			\textit{modify} method arguments           & 7  & 4    & 7 \\ 
			\textit{add} \J{null} check                     & 1 & 16     & 3 \\ 
			\textit{modify} condition (strengthen)          & 7  & 3    & 14  \\ 
			\textit{add} field initialization $+$ \textit{modify} method signature & 6  & 2    & 2 \\ 
			\textit{modify} condition (other)              & 3 & 1    & 2 \\ 
			\textit{add} field initialization $+$ \textit{modify} method call arguments   & 2 & 1    & 2 \\ 
			\textit{add} method call $+$ \textit{add} method signature             & 8  & 5     & 8  \\ 
			\textit{modify} assignment expression       & 2  & 2    & 3 \\ 
			\textit{add} conditional $+$ \textit{modify} block scope   & 1 & 7    & 1 \\ \bottomrule
		\end{tabular}%
	}
\end{table}

\paragraph{Patch classification}
Based on a systematic analysis of the selected \numprint{6583} bug-fixing commits,
we identified 67 individual constructs;
\Cref{tab:patch-kinds} lists the most common ones.
Since a patch may combine any number of these constructs,
there is an astronomically large number
of $2^{67} \simeq 10^{20}$ possible unique patch kinds;
in practice, though, we only found a much smaller number of
\numprint{229} patch kinds.
This indicates that ``real'' patches
usually only involve a small number of constructs,
and combine them by predictable patterns%
---which is consistent with the generally observed
``naturalness'' of software~\cite{naturalness2012,naturalness-buggy}.
\Cref{tab:category} shows the ten most
frequently observed patch kinds,
and the corresponding bugs they fix.

\paragraph{Historic frequencies}
The last step of \tool's training phase is the computation
of historic frequencies $\freq[c][k]$
for each bug category $c$ and fix kind~$k$.
This step is computationally trivial, as it only
requires to compute frequencies in a dataset.
In our experiments, \tool took less than 1~minute
to compute the frequencies for all selected \numprint{6583}
bug-fixing commits.

\subsection{Subjects: APR Tools and Patches}
\label{sec:subjects}

The overall goal of this evaluation is
assessing whether \tool can \emph{improve} the ranking of correct
patches (over merely plausible ones)
compared to that produced by the APR tools that produced the patches in the first place.
In order to also demonstrate flexibility,
the evaluation should include as many different APR tools---and as many different patches---as possible.

\begin{table}
	\resizebox{\columnwidth}{!}{%
		\begin{tabular}{cl *{4}{c}}
        \toprule
        \multicolumn{1}{c}{\textsc{}}
        & \multicolumn{1}{c}{\textsc{tools}}
        & \multicolumn{1}{c}{\textsc{av?}}
        & \multicolumn{1}{c}{$>1$\textsc{?}}
        & \multicolumn{1}{c}{\textsc{no pfl?}}
        & \multicolumn{1}{c}{$\geq10$\textsc{?}}
        \\ \midrule
			\multirow{2}{*}{\textsl{analysis}}         & SimFix~\cite{jiang2018shaping}, kPAR~\cite{liu2019you}, (jGenProg, jKali, jMutRepair)~\cite{martinez2016astor}, DeepRepair~\cite{DeepRepair}                                                                       & \includeOK      & \includeOK &  \includeOK & \includeNO \\ \cmidrule(l){2-6} 
        & \begin{tabular}[c]{@{}l@{}}ARJA~\cite{arja}, Cardumen~\cite{cardumen}, Jaid~\cite{pei_jaid_2017}, RSRepair~\cite{rsrepair}, TBar~\cite{tbar}\end{tabular}               & \includeOK       & \includeOK &  \includeOK &  \includeOK  \\
        \midrule
        \multirow{5}{*}{\textsl{learning}} %
		
			& Gamma~\cite{Gamma2023}, Circle~\cite{yuan2022circle}, Cure~\cite{jiang2021cure}, SEQUENCER~\cite{chen2019sequencer}         & \includeOK & \includeOK & \includeNO  \\ \cmidrule(l){2-6} 
			& Dlfix~\cite{li2020dlfix}, KNOD~\cite{jiang2023knod}, Repilot~\cite{wei2023copiloting}																				& \includeOK & \includeNO &  \\ \cmidrule(l){2-6}
			& AlphaRepair~\cite{xia2022less},  CoCoNuT~\cite{lutellier2020coconut}, TRANSFER~\cite{Transfer}																					& \includeNO \\ \cmidrule(l){2-6}
			& Recoder~\cite{zhu2021syntax}, Tenure~\cite{Tenure2023}  																																& \includeOK & \includeNO \\ \cmidrule(l){2-6} 
			& ITER~\cite{ye2024iter}, RapidCapr~\cite{hidvegi2023token}, RewardRepair~\cite{ye2022neural}                                                                                       & \includeOK       & \includeOK &  \includeOK &  \includeOK \\ \bottomrule
		\end{tabular}%
	}
	\caption{Automated program repair tool selection.
     For every group of APR \textsc{tools},
     whether their experimental artifacts (patches) are available for Defects4J bugs (column \textsc{av?}),
     whether they include more than one plausible patch per bug ($>1$\textsc{?}),
     whether they did \emph{not} use perfect fault localization (\textsc{no pfl?}),
     and whether they include at least 5 plausible patches
     for at least ten fixed Defects4J bugs ($\geq 10$\textsc{?}).}
	\label{tab:tool-selection}
\end{table}

\begin{table*}
  \setlength{\tabcolsep}{4pt}
  \centering
  \small
 \resizebox{\textwidth}{!}{ \begin{tabular}{l *{6}{rrr} |rrr}
    \toprule
    & \multicolumn{3}{c}{\textsl{Chart}}
    & \multicolumn{3}{c}{\textsl{Closure}}
    & \multicolumn{3}{c}{\textsl{Lang}}
    & \multicolumn{3}{c}{\textsl{Math}}
    & \multicolumn{3}{c}{\textsl{Mockito}}
    & \multicolumn{3}{c|}{\textsl{Time}}
    & \multicolumn{3}{c}{\textsl{all bugs}}
    \\
    \cmidrule(lr){2-4}
    \cmidrule(lr){5-7}
    \cmidrule(lr){8-10}
    \cmidrule(lr){11-13}
    \cmidrule(lr){14-16}
    \cmidrule(lr){17-19}
    \cmidrule(lr){20-22}
    \multicolumn{1}{c}{\textsc{tool}}
    & \multicolumn{1}{c}{\textsc{f}}
    & \multicolumn{1}{c}{\textsc{p}}
    & \multicolumn{1}{c}{\textsc{b}}
    & \multicolumn{1}{c}{\textsc{f}}
    & \multicolumn{1}{c}{\textsc{p}}
    & \multicolumn{1}{c}{\textsc{b}}
    & \multicolumn{1}{c}{\textsc{f}}
    & \multicolumn{1}{c}{\textsc{p}}
    & \multicolumn{1}{c}{\textsc{b}}
    & \multicolumn{1}{c}{\textsc{f}}
    & \multicolumn{1}{c}{\textsc{p}}
    & \multicolumn{1}{c}{\textsc{b}}
    & \multicolumn{1}{c}{\textsc{f}}
    & \multicolumn{1}{c}{\textsc{p}}
    & \multicolumn{1}{c}{\textsc{b}}
    & \multicolumn{1}{c}{\textsc{f}}
    & \multicolumn{1}{c}{\textsc{p}}
    & \multicolumn{1}{c|}{\textsc{b}}
    & \multicolumn{1}{c}{\textsc{f}}
    & \multicolumn{1}{c}{\textsc{p}}
    & \multicolumn{1}{c}{\textsc{b}}
    \\
    \midrule
    ARJA & 4 & 23 & 6 &   &   &   & 7 & 38 & 5 & 13 & 90 & 7 &   &   &   & 3 & 24 & 8 & 27 & 175 & 6 
\\
Cardumen & 6 & 30 & 5 & 3 & 15 & 5 & 5 & 25 & 5 & 24 & 120 & 5 &   &   &   & 3 & 15 & 5 & 41 & 205 & 5 
\\
ITER & 2 & 62 & 31 & 10 & 130 & 13 & 3 & 30 & 10 & 14 & 137 & 10 &   &   &   & 1 & 24 & 24 & 30 & 383 & 13 
\\
Jaid & 7 & 1918 & 274 & 19 & 2167 & 114 & 16 & 1347 & 84 & 20 & 2803 & 140 & 2 & 23 & 12 & 2 & 66 & 33 & 66 & 8324 & 126 
\\
RSRepair & 4 & 29 & 7 &   &   &   & 5 & 37 & 7 & 12 & 199 & 17 &   &   &   & 2 & 30 & 15 & 23 & 295 & 13 
\\
RapidCapr & 7 & 348 & 50 & 24 & 625 & 26 & 13 & 383 & 29 & 23 & 721 & 31 & 5 & 130 & 26 & 3 & 64 & 21 & 75 & 2271 & 30 
\\
RewardRepair & 2 & 377 & 188 & 23 & 4588 & 199 & 11 & 2200 & 200 & 10 & 1999 & 200 & 3 & 600 & 200 & 2 & 400 & 200 & 51 & 10164 & 199 
\\
T-BAR & 4 & 112 & 28 & 8 & 338 & 42 & 7 & 150 & 21 & 12 & 580 & 48 &   &   &   & 2 & 35 & 18 & 33 & 1215 & 37 
\\
\cmidrule(lr){2-22}
\textsl{all tools} & 15 & 2899 & 193 & 50 & 7863 & 157 & 32 & 4210 & 132 & 54 & 6649 & 123 & 9 & 753 & 84 & 8 & 658 & 82 & 168 & 23032 & 137

    \\
    \bottomrule
  \end{tabular}}
  \caption{Summary of the patches used in \tool's experiments.
    For each APR \textsc{tool}, for each Defects4J project identifier
    (\textsl{Chart}, \textsl{Closure}, \ldots),
    the table reports:
    the number \textsc{f} of faults of that project correctly fixed by the tool;
    the total number \textsc{p} of plausible patches produced by the tool for those faults;
    the average number \textsc{b} of patches produced by the tool for each of those faults (i.e., $\textsc{b} = \textsc{p}/\textsc{f}$).
    The bottom row combines data from \textsl{all tools};
    the rightmost column combines data from all Defects4J projects.
    }
  \label{tab:patches-subjects}
\end{table*}

\paragraph{APR tool selection}
To this end,
we initially considered the 26 APR tools listed in \Cref{tab:tool-selection}:
this list includes implementations of traditional approaches
(e.g., jGenProg, kPAR),
more advanced approaches combining different dynamic analysis techniques
(e.g., Jaid, TBar),
as well as numerous recent approaches that use some form of machine learning
(e.g., AlphaRepair, ITER)---which have grown in popularity in recent years.

We culled this initial list by dropping any APR tool
that does not satisfy all these criteria:
\begin{enumerate*}
\item The tool has been evaluated on the Defects4J benchmark (v.~1.2.0 or later~\cite{defects4j, just2014defects4j}),
  and its experimental artifacts
  (in particular, the patches generated in the experiments)
  are publicly available.
  This criterion is obviously needed to ensure that we can feed the tool's patches to \tool.
\item The tool's patches include multiple patches for the same bug.
  If a tool produces at most one patch per bug, there is nothing to rank, and hence \tool's capabilities are irrelevant.
  Importantly, even if an APR tool could, in principle, produce multiple patches per bug,
  we do not consider it if it has not been evaluated in this scenario in the tool's paper's evaluation;
  in other words, we only consider the tool's performance in the experimental conditions set by its authors.\iflong\footnote{
      For example: AlphaRepair does not include patches but its implementation is available. We ran it (settings: Ochai fault localization, top-40 most suspicious locations, beam setting 5) on Defects4J bugs, but it generated more than 4 plausible patches for only 3 bugs. This does not mean that it is not capable of producing more patches with different settings; however, we stick to the experimental artifacts provided by its authors, which are more likely to demonstrate an optimal usage of the tool.}\fi
  As we discuss in \Cref{sec:related-work},
  this criterion also excludes most recent LLM-based APR approaches,
  which typically employ an iterative patch generation process,
  producing only a very small number of candidates per iteration.
  
\item The tool does \emph{not} use perfect fault localization
  (i.e., it only produces patches that modify the location of the programmer-written fix, which is given as input).
  Again, even if a tool \emph{could} take a different fault localization input,
  we stick to the way it has been evaluated by its authors.
  The reason for avoiding patches generated with perfect fault localization is that, even if there are multiple patches per bug,
  they all tend to be very similarly syntactically, since they all originate from the same key information about the bug location;
  hence, they have little discriminatory power for a feature-based approach like \tool's
  (which, in fact, would tend to merely replicate the same ranking as the original tool).
\item Finally, the APR tool's available patches should consist of at least 5 patches for each of at least 10 Defects4J bugs that it can fix correctly.
  This criterion is simply to ensure that we have a nontrivial collection of subjects to experiment with \tool
  used on a certain APR tool's output.
\end{enumerate*}
\Cref{tab:tool-selection} details which tools satisfy each of these criteria.
We ended up with a selection of 8 APR tools
(ARJA, Cardumen, ITER, Jaid, RapidCapr, RewardRepair, RSRepair, and TBar)
that satisfy all criteria.

\paragraph{Patch selection}
From the experimental artifacts of the 8 selected tools,
we collected all available plausible patches for Defects4J bugs
such that the tool produces at least one correct fix
and at least four plausible patches.
If a tool only produces plausible, incorrect patches for a bug,
their ranking is immaterial. Also, if a tool produces fewer than
four plausible patches for a bug, even the worst ranking will
put a correct fix in the top-3; hence, we exclude these bugs
from our evaluation as they would not be relevant to assess effectiveness.
We use the widely accepted definition of correctness:
a patch is a correct fix if it is
semantically equivalent to the programmer-written fix for the same bug
(which is available in Defects4J).
\Cref{tab:patches-subjects}
gives some details about the selected bugs and patches.
Overall, the 8 APR tools included correct fixes
(and at least four plausible patches)
for 168 bugs (123 logic, 34 null pointer, 11 overflow),
and \numprint{23032} plausible patches
(\numprint{15601} logic, \numprint{5272} null pointer, \numprint{2159} overflow).
This selection of patches is quite varied,
as it includes patches of bugs from 6 different Defects4J projects,
and tools that produce a range of different plausible patches
(from Cardumen's 5 plausible patches per bug,
up to RewardRepair's whopping 199 per bug).

\begin{table*}
  \setlength{\tabcolsep}{2pt}
  \centering
  \footnotesize
\resizebox{\textwidth}{!}{  \begin{tabular}{l rr *{8}{r} *{8}{r} *{8}{r} *{8}{r}}
    \toprule
    & &
    & \multicolumn{8}{c}{\textsc{top 1}}
    & \multicolumn{8}{c}{\textsc{top 3}}
    & \multicolumn{8}{c}{\textsc{top 5}}
    & \multicolumn{8}{c}{\textsc{top 10}}
    \\
    \cmidrule(lr){4-11}
    \cmidrule(lr){12-19}
    \cmidrule(lr){20-27}
    \cmidrule(lr){28-35}
    & &
    & \multicolumn{4}{c}{\textsc{\#}}
    & \multicolumn{4}{c}{\textsc{\%}}
    & \multicolumn{4}{c}{\textsc{\#}}
    & \multicolumn{4}{c}{\textsc{\%}}
    & \multicolumn{4}{c}{\textsc{\#}}
    & \multicolumn{4}{c}{\textsc{\%}}
    & \multicolumn{4}{c}{\textsc{\#}}
    & \multicolumn{4}{c}{\textsc{\%}}
    \\
    \cmidrule(lr){4-7}
    \cmidrule(lr){8-11}
    \cmidrule(lr){12-15}
    \cmidrule(lr){16-19}
    \cmidrule(lr){20-23}
    \cmidrule(lr){24-27}
    \cmidrule(lr){28-31}
    \cmidrule(lr){32-35}
    \multicolumn{1}{c}{\textsc{tool}}
    & \multicolumn{1}{c}{\textsc{k}}
    & \multicolumn{1}{c}{\textsc{bugs}}
    & \multicolumn{1}{c}{\textsc{n}}
    & \multicolumn{1}{c}{\textsc{b}}
    & \multicolumn{1}{c}{\textsc{o}}
    & \multicolumn{1}{c}{\textsc{p}}
    & \multicolumn{1}{c}{\textsc{n}}
    & \multicolumn{1}{c}{\textsc{b}}
    & \multicolumn{1}{c}{\textsc{o}}
    & \multicolumn{1}{c}{\textsc{p}}
    & \multicolumn{1}{c}{\textsc{n}}
    & \multicolumn{1}{c}{\textsc{b}}
    & \multicolumn{1}{c}{\textsc{o}}
    & \multicolumn{1}{c}{\textsc{p}}
    & \multicolumn{1}{c}{\textsc{n}}
    & \multicolumn{1}{c}{\textsc{b}}
    & \multicolumn{1}{c}{\textsc{o}}
    & \multicolumn{1}{c}{\textsc{p}}
    & \multicolumn{1}{c}{\textsc{n}}
    & \multicolumn{1}{c}{\textsc{b}}
    & \multicolumn{1}{c}{\textsc{o}}
    & \multicolumn{1}{c}{\textsc{p}}
    & \multicolumn{1}{c}{\textsc{n}}
    & \multicolumn{1}{c}{\textsc{b}}
    & \multicolumn{1}{c}{\textsc{o}}
    & \multicolumn{1}{c}{\textsc{p}}
    & \multicolumn{1}{c}{\textsc{n}}
    & \multicolumn{1}{c}{\textsc{b}}
    & \multicolumn{1}{c}{\textsc{o}}
    & \multicolumn{1}{c}{\textsc{p}}
    & \multicolumn{1}{c}{\textsc{n}}
    & \multicolumn{1}{c}{\textsc{b}}
    & \multicolumn{1}{c}{\textsc{o}}
    & \multicolumn{1}{c}{\textsc{p}}
    \\
    \midrule
ARJA & \textsc{*} & 27 & 16 & 6 & 1 & \best{4} & 59{\tiny\%} & 22{\tiny\%} & 4{\tiny\%} & \best{15{\tiny\%}} & 5 & 16 & 0 & \best{6} & 19{\tiny\%} & 59{\tiny\%} & 0{\tiny\%} & \best{22{\tiny\%}} & 1 & 24 & 0 & \best{2} & 4{\tiny\%} & 89{\tiny\%} & 0{\tiny\%} & \best{7{\tiny\%}} & 0 & 27 & 0 & 0 & 0{\tiny\%} & 100{\tiny\%} & 0{\tiny\%} & 0{\tiny\%}
\\
Cardumen & \textsc{*} & 41 & 18 & 9 & 4 & \best{10} & 44{\tiny\%} & 22{\tiny\%} & 10{\tiny\%} & \best{24{\tiny\%}} & 1 & 28 & 0 & \best{12} & 2{\tiny\%} & 68{\tiny\%} & 0{\tiny\%} & \best{29{\tiny\%}} & 0 & 41 & 0 & 0 & 0{\tiny\%} & 100{\tiny\%} & 0{\tiny\%} & 0{\tiny\%} & 0 & 41 & 0 & 0 & 0{\tiny\%} & 100{\tiny\%} & 0{\tiny\%} & 0{\tiny\%}
\\
ITER & \textsc{*} & 30 & 17 & 6 & 2 & \best{5} & 57{\tiny\%} & 20{\tiny\%} & 7{\tiny\%} & \best{17{\tiny\%}} & 3 & 15 & 1 & \best{11} & 10{\tiny\%} & 50{\tiny\%} & 3{\tiny\%} & \best{37{\tiny\%}} & 0 & 23 & 1 & \best{6} & 0{\tiny\%} & 77{\tiny\%} & 3{\tiny\%} & \best{20{\tiny\%}} & 0 & 26 & 0 & \best{4} & 0{\tiny\%} & 87{\tiny\%} & 0{\tiny\%} & \best{13{\tiny\%}}
\\
Jaid & \textsc{*} & 66 & 38 & 16 & 5 & \best{7} & 58{\tiny\%} & 24{\tiny\%} & 8{\tiny\%} & \best{11{\tiny\%}} & 22 & 30 & 0 & \best{14} & 33{\tiny\%} & 45{\tiny\%} & 0{\tiny\%} & \best{21{\tiny\%}} & 12 & 35 & 2 & \best{17} & 18{\tiny\%} & 53{\tiny\%} & 3{\tiny\%} & \best{26{\tiny\%}} & 6 & 45 & 1 & \best{14} & 9{\tiny\%} & 68{\tiny\%} & 2{\tiny\%} & \best{21{\tiny\%}}
\\
RapidCapr & \textsc{*} & 75 & 52 & 14 & 4 & \best{5} & 69{\tiny\%} & 19{\tiny\%} & 5{\tiny\%} & \best{7{\tiny\%}} & 25 & 31 & 3 & \best{16} & 33{\tiny\%} & 41{\tiny\%} & 4{\tiny\%} & \best{21{\tiny\%}} & 18 & 39 & 3 & \best{15} & 24{\tiny\%} & 52{\tiny\%} & 4{\tiny\%} & \best{20{\tiny\%}} & 9 & 58 & 0 & \best{8} & 12{\tiny\%} & 77{\tiny\%} & 0{\tiny\%} & \best{11{\tiny\%}}
\\
RewardRepair & \textsc{*} & 51 & 30 & 8 & 4 & \best{9} & 59{\tiny\%} & 16{\tiny\%} & 8{\tiny\%} & \best{18{\tiny\%}} & 5 & 18 & 1 & \best{27} & 10{\tiny\%} & 35{\tiny\%} & 2{\tiny\%} & \best{53{\tiny\%}} & 0 & 28 & 0 & \best{23} & 0{\tiny\%} & 55{\tiny\%} & 0{\tiny\%} & \best{45{\tiny\%}} & 0 & 39 & 0 & \best{12} & 0{\tiny\%} & 76{\tiny\%} & 0{\tiny\%} & \best{24{\tiny\%}}
\\
RSRepair & \textsc{*} & 23 & 16 & 1 & 1 & \best{5} & 70{\tiny\%} & 4{\tiny\%} & 4{\tiny\%} & \best{22{\tiny\%}} & 5 & 10 & 0 & \best{8} & 22{\tiny\%} & 43{\tiny\%} & 0{\tiny\%} & \best{35{\tiny\%}} & 2 & 15 & 0 & \best{6} & 9{\tiny\%} & 65{\tiny\%} & 0{\tiny\%} & \best{26{\tiny\%}} & 2 & 19 & 0 & \best{2} & 9{\tiny\%} & 83{\tiny\%} & 0{\tiny\%} & \best{9{\tiny\%}}
\\
T-BAR & \textsc{*} & 33 & 15 & 14 & 1 & \best{3} & 45{\tiny\%} & 42{\tiny\%} & 3{\tiny\%} & \best{9{\tiny\%}} & 4 & 21 & 2 & \best{6} & 12{\tiny\%} & 64{\tiny\%} & 6{\tiny\%} & \best{18{\tiny\%}} & 2 & 25 & 2 & \best{4} & 6{\tiny\%} & 76{\tiny\%} & 6{\tiny\%} & \best{12{\tiny\%}} & 1 & 29 & 1 & \best{2} & 3{\tiny\%} & 88{\tiny\%} & 3{\tiny\%} & \best{6{\tiny\%}}
\\
\midrule
\multirow{4}{*}{\textsl{all tools}} & \textsc{l} & 236 & 135 & 56 & 13 & \best{32} & 57{\tiny\%} & 24{\tiny\%} & 6{\tiny\%} & \best{14{\tiny\%}} & 46 & 111 & 6 & \best{73} & 19{\tiny\%} & 47{\tiny\%} & 3{\tiny\%} & \best{31{\tiny\%}} & 22 & 153 & 8 & \best{53} & 9{\tiny\%} & 65{\tiny\%} & 3{\tiny\%} & \best{22{\tiny\%}} & 11 & 194 & 1 & \best{30} & 5{\tiny\%} & 82{\tiny\%} & 0{\tiny\%} & \best{13{\tiny\%}}
\\
 & \textsc{n} & 83 & 49 & 17 & 6 & \best{11} & 59{\tiny\%} & 20{\tiny\%} & 7{\tiny\%} & \best{13{\tiny\%}} & 17 & 46 & 0 & \best{20} & 20{\tiny\%} & 55{\tiny\%} & 0{\tiny\%} & \best{24{\tiny\%}} & 7 & 60 & 0 & \best{16} & 8{\tiny\%} & 72{\tiny\%} & 0{\tiny\%} & \best{19{\tiny\%}} & 6 & 68 & 0 & \best{9} & 7{\tiny\%} & 82{\tiny\%} & 0{\tiny\%} & \best{11{\tiny\%}}
\\
 & \textsc{o} & 27 & 18 & 1 & 3 & \best{5} & 67{\tiny\%} & 4{\tiny\%} & 11{\tiny\%} & \best{19{\tiny\%}} & 7 & 12 & 1 & \best{7} & 26{\tiny\%} & 44{\tiny\%} & 4{\tiny\%} & \best{26{\tiny\%}} & 6 & 17 & 0 & \best{4} & 22{\tiny\%} & 63{\tiny\%} & 0{\tiny\%} & \best{15{\tiny\%}} & 1 & 22 & 1 & \best{3} & 4{\tiny\%} & 81{\tiny\%} & 4{\tiny\%} & \best{11{\tiny\%}}
\\
 & \textsc{*} & 346 & 202 & 74 & 22 & \best{48} & 58{\tiny\%} & 21{\tiny\%} & 6{\tiny\%} & \best{14{\tiny\%}} & 70 & 169 & 7 & \best{100} & 20{\tiny\%} & 49{\tiny\%} & 2{\tiny\%} & \best{29{\tiny\%}} & 35 & 230 & 8 & \best{73} & 10{\tiny\%} & 66{\tiny\%} & 2{\tiny\%} & \best{21{\tiny\%}} & 18 & 284 & 2 & \best{42} & 5{\tiny\%} & 82{\tiny\%} & 1{\tiny\%} & \best{12{\tiny\%}}
    \\ \bottomrule
  \end{tabular}}
  \caption{Comparison between \tool and other APR tools.
    Each row reports the number of \textsc{bugs} of category \textsc{k} for which the given APR \textsc{tool} produces at least
    one correct fix; \textsc{*} denotes bugs of any category,
    whereas \textsc{l} are logical, \textsc{n} are null-pointer,
    and \textsc{o} are overflow bugs.
    The internal columns indicate the number \textsc{\#} and percentage \textsc{\%} of these bugs \textsc{bugs}
    that fall into one of four scenarios:
    \textsc{n}) neither the other tool nor \tool, rank a correct fix in the
    \textsc{top-$n$};
    \textsc{b}) both the other tool and \tool rank a correct fix in the \textsc{top-$n$};
    \textsc{o}) only the other tool ranks a correct fix in the \textsc{top-$n$}
    (\tool does not);
    \textsc{p}) only \tool ranks a correct fix in the \textsc{top-$n$}
    (the other tool does not).
    The bottom part of the table considers all tools together,
    with a breakdown by bug category (\textsc{l}ogical, \textsc{n}ull, and \textsc{o}verflow).
    All scenarios where \tool strictly improves over the other tool
    are \best{highlighted}.
  }
  \label{tab:rq1-main}
\end{table*}

\begin{table}
  \setlength{\tabcolsep}{4pt}
  \centering
  \footnotesize
  \begin{tabular}{l l *{4}{rr}}
    \toprule
    &
    & \multicolumn{2}{c}{\textsc{min}}
    & \multicolumn{2}{c}{\textsc{average}}
    & \multicolumn{2}{c}{\textsc{median}}
    & \multicolumn{2}{c}{\textsc{max}}
    \\
    \cmidrule(lr){3-4}
    \cmidrule(lr){5-6}
    \cmidrule(lr){7-8}
    \cmidrule(lr){9-10}
    \multicolumn{1}{c}{\textsc{category}}
    & \multicolumn{1}{c}{\textsc{tool}}
    & \multicolumn{1}{c}{\textsc{o}} & \multicolumn{1}{c}{\textsc{p}}
    & \multicolumn{1}{c}{\textsc{o}} & \multicolumn{1}{c}{\textsc{p}}
    & \multicolumn{1}{c}{\textsc{o}} & \multicolumn{1}{c}{\textsc{p}}
    & \multicolumn{1}{c}{\textsc{o}} & \multicolumn{1}{c}{\textsc{p}}
    \\
    \midrule
\multirow{9}{*}{\textsl{any bug}} & ARJA & 1 & 1 & 3.2 & 2.5 & 3.0 & 2.0 & 10 & 8 \\
  & Cardumen & 1 & 1 & 2.5 & 1.8 & 2.0 & 2.0 & 5 & 5 \\
  & ITER & 1 & 1 & 5.6 & 2.2 & 3.0 & 2.0 & 30 & 6 \\
  & Jaid & 1 & 1 & 27.0 & 7.0 & 4.0 & 2.0 & 789 & 119 \\
  & RSRepair & 1 & 1 & 5.5 & 3.4 & 4.0 & 2.0 & 14 & 18 \\
  & RapidCapr & 1 & 1 & 8.1 & 5.4 & 4.0 & 3.0 & 80 & 34 \\
  & RewardRepair & 1 & 1 & 17.7 & 2.1 & 5.0 & 2.0 & 162 & 5 \\
  & T-BAR & 1 & 1 & 6.7 & 4.3 & 3.0 & 1.0 & 106 & 45 \\
\cmidrule(lr){3-10}
 & \textsl{all tools} & 1 & 1 & 11.5 & 4.1 & 3.0 & 2.0 & 789 & 119
    \\ \bottomrule
  \end{tabular}
  \caption{Statistics about ranks of correct fixes comparing \tool to other APR tools.
      For
    every \textsc{tool} (and \emph{all tools} together),
    the table reports the \textsc{min}imum, \textsc{average} (mean), \textsc{median}, and \textsc{max}imum rank of the correct fix in the tool's original ranking
  (columns \textsc{o}) and in \tool's ranking (columns \textsc{p}).}
  \label{tab:rq1-stats}
\end{table}

\begin{figure}
  \centering
  \includegraphics[width=0.75\textwidth]{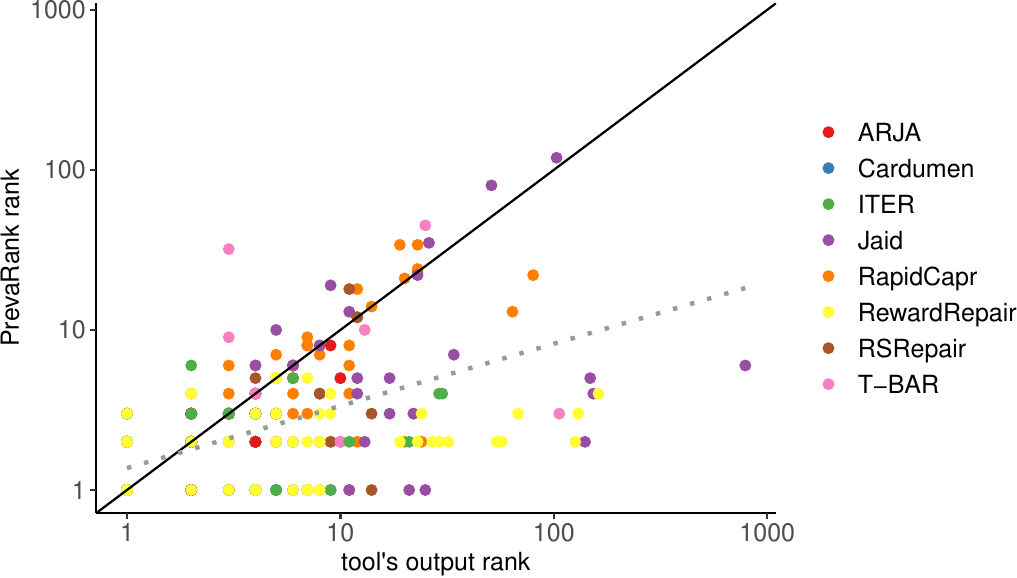}
  \caption{Each point represents a bug fixed by an APR tool. The point's $x$ coordinate is the rank of the (first) correct fix assigned by the APR tool; the point's $y$ coordinate is the rank assigned by \tool.
    Thus, points below the diagonal line correspond to bugs where \tool improved the APR tool's ranking of correct fixes. The dotted line is the linear regression line of the points, which highlights the data trend. Axis scales are logarithmic.}
  \label{fig:scatterplot-ranks}
\end{figure}

\subsection{RQ1: Effectiveness}
\label{sec:rq1}

We conducted two series of experiments to evaluate
\tool's effectiveness.
In the first series,
each run of \tool ranks the plausible patches $P(t, b)$
produced by APR tool $t$ for bug $b$.
In the second series,
each run of \tool ranks the plausible patches $P(b)$
produced by any APR tools that could fix bug $b$.
In other words, in the first series
each tool is considered \emph{individually},
whereas in the second series patches from different tools
are ranked together.

\paragraph{Individual tool patch ranking}
Each point in \Cref{fig:scatterplot-ranks}'s scatterplot
represents one of the 168 bugs used in the experiments;
a point at coordinates $(x, y)$
denotes that the APR tool ranked at position $x$ the (first) correct fix
for the corresponding bug, whereas \tool ranked it at position $y$.
Visually, it is clear that the majority of points are below
the diagonal $x = y$ line (i.e., $y < x$), which means that \tool
usually improves (lower is better) the ranking of correct fixes.
On the other hand,
these points do not seem to follow any distinct pattern
per APR tool (denoted by their color),
which suggests that \tool tends to work well
regardless of which APR tool produced the patches it is ranking.

\begin{figure*}
  \centering
  \includegraphics[width=\textwidth]{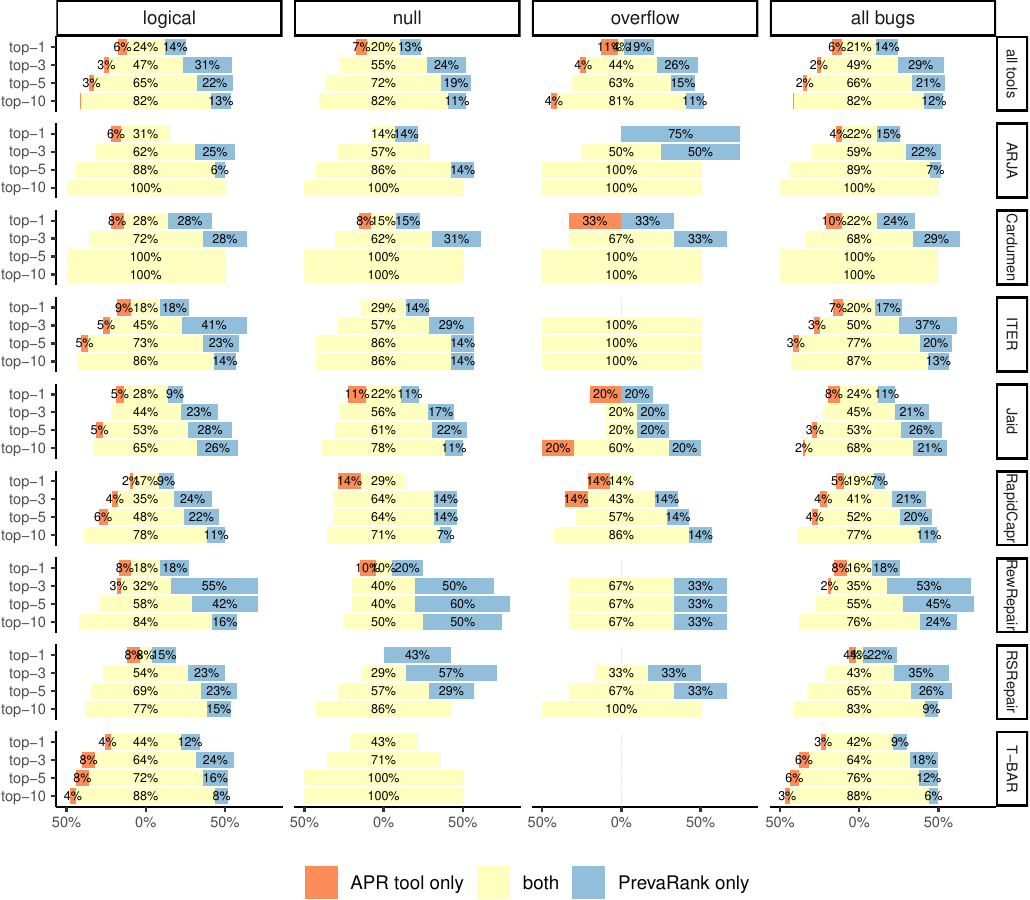}
	\caption{Percentage of bugs of each category, among those for which each APR tool can generate at least one correct fix, that are ranked in the top-1, top-3, top-5, and top-10 positions only in the APR tool's original ranking, only in \tool's ranking, or in both rankings. (These groupings correspond to \cref{tab:rq1-main}'s three scenarios \textsc{o}, \textsc{p}, and \textsc{b}, respectively.)}
	\label{fig:percent_topk}
\end{figure*}

\Cref{tab:rq1-main}
details the effectiveness of \tool
quantitatively:
for each APR tool,
it partitions the bugs fixed by that tool
in 4 groups:
group \textsc{n} includes
those for which a correct fix appears
within the top-$k$
neither in the tool's original ranking
nor in \tool's ranking;
group \textsc{b} includes
those for which a correct fix appears
within the top-$k$ in both the tool's and \tool's rankings;
group \textsc{o} includes
those for which a correct fix appears within the top-$k$
only in the tool's original ranking;
and group \textsc{p} includes 
those for which a correct fix appears within the top-$k$
only in \tool's ranking.
Equivalently, group \textsc{p} consists of
the bugs where \tool \emph{improves} the ranking of correct fixes
from outside to inside the top-$k$;
and group \textsc{o} consists of
the bugs where \tool \emph{worsens} the ranking of correct fixes
from inside to outside the top-$k$.
As it is customary, we consider the top-1, top-3, top-5, and top-10
positions in the ranking,
which reflect realistic scenarios~\cite{trust,trust2,meem2024exploring}
where a developer would be willing to inspect at most ten fixes
to identify a correct one.
\Cref{fig:percent_topk}
displays similar data as \Cref{tab:rq1-main}
with an additional break-down by bug category. %

Within this data,
it is interesting to compare the size of groups \textsc{p} and \textsc{o},
which gives an idea \tool's ranking quality
compared to the original APR tool's.
The first thing we notice is that
the size of group \textsc{p} is practically always
greater than the size of group \textsc{o};
the only exceptions are when both the tool's original ranking
and \tool's ranking include all correct fixes within the top-$k$.
This indicates that, while \tool's ranking is not infallible
(as there are situations where the original APR tool's ranking is more accurate),
it systematically provides a net improvement.

In fact, the size of group \textsc{p} is not just barely bigger than
the size of group \textsc{o}:
in 87\% of cases, it is at least twice as big;
in 23\% of cases, it is 10 times larger or more.
This is easy to see in \Cref{fig:percent_topk},
where \tool's bar segments are often several times wider than the original tool's.
Thus, \tool brings a quantitative advantage
with its capability of improving the ranking of numerous correct fixes.
We also see that \tool's effectiveness is largely independent of
the category of bugs, APR tools, or top-$k$ rank we consider.
The few outliers usually correspond to narrow, specific scenarios.
For instance, Jaid ranks in position 1 the correct fix for
an overflow bug, which \tool ranks in position 2;
still, \tool ranks in position~1 the correct fix for another
overflow bug, which Jaid ranks in position 7.
Another example are the correct fixes for three overflow bugs
that Cardumen ranks at positions 1, 2, and 4,
whereas \tool ranks at positions 3, 3, and 1.
In all, we are talking about small differences in rank
in a few corner cases.

Another trend visible in \Cref{tab:rq1-main} is that
the difference between groups \textsc{p} and \textsc{o} does
not grow monotonically with the rank cutoff $k$ but peaks for top-3.
This behavior is simply a saturation effect: there aren't many bugs
that admit many different plausible patches.

\Cref{tab:rq1-stats} gives a final view on the effectiveness
of \tool in ranking the patches produced by each APR tool.
The average (mean and median) rank of a correct fix in
\tool's ranking is nearly always better (i.e., lower position)
than the rank in the APR tool's original ranking;
\tool's overall mean rank is nearly $1/3$ of the original rank
(4.1 vs.~11.5).
It is also interesting that \tool often improves the ranking in
worst-case scenarios where there are a very large number of plausible, incorrect
patches;
the most extreme case is an overflow bug for which Jaid produces
810 plausible patches:
Jaid ranks the correct fix at position 789, whereas \tool massively improves
its rank to position~6.

\begin{table}
  \centering
  \setlength{\tabcolsep}{2.5pt}
  \footnotesize
  \begin{tabular}{c *{5}{rr}}
    \toprule
    & \multicolumn{2}{c}{\textsc{rank}}
    & \multicolumn{2}{c}{\textsc{top 1}}
    & \multicolumn{2}{c}{\textsc{top 3}}
    & \multicolumn{2}{c}{\textsc{top 5}}
    & \multicolumn{2}{c}{\textsc{top 10}}
    \\
    \cmidrule(lr){2-3}
    \cmidrule(lr){4-5}
    \cmidrule(lr){6-7}
    \cmidrule(lr){8-9}
    \cmidrule(lr){10-11}
    & \multicolumn{1}{c}{\textsc{a}}
    & \multicolumn{1}{c}{\textsc{m}}
    & \multicolumn{1}{c}{\textsc{\#}}
    & \multicolumn{1}{c}{\textsc{\%}}
    & \multicolumn{1}{c}{\textsc{\#}}
    & \multicolumn{1}{c}{\textsc{\%}}
    & \multicolumn{1}{c}{\textsc{\#}}
    & \multicolumn{1}{c}{\textsc{\%}}
    & \multicolumn{1}{c}{\textsc{\#}}
    & \multicolumn{1}{c}{\textsc{\%}}
    \\
    \midrule
    \textsl{APR tools} & 2.1 & 1.0 & 57 & 61{\small \%} & 79 & 85{\small \%} & 87 & 94{\small \%} & 92 & 99{\small \%}
\\
\textsl{\tool individual} & 1.4 & 1.0 & 63 & 68{\small \%} & 91 & 98{\small \%} & 93 & 100{\small \%} & 93 & 100{\small \%}
\\
\textsl{\tool cumulative} & 1.6 & 1.0 & 48 & 52{\small \%} & 93 & 100{\small \%} & 93 & 100{\small \%} & 93 & 100{\small \%}
    \\
    \bottomrule
  \end{tabular}
  \caption{Effectiveness of \tool when ranking multiple tool patches.
    The table compares the best (highest) rank
    of all \textsl{APR tools},
    to the best \tool rank working
    on each tool's patches \textsl{individually},
    to \tool rank's on all patches \textsl{cumulatively}.
    It reports the \textsc{a}verage (mean) and \textsc{m}edian rank,
    and the number \textsc{\#} and percentage \textsc{\%}
    (over all 93 bugs that more than one tool fixed) of bugs
  in the \textsc{top $k$}.}
	\label{tab:cumulative}
\end{table}

\paragraph{Multiple tool patch ranking}
Let us now consider only the 93 Defects4J bugs
(63 logical, 24 null pointer, 6 overflow)
that at least two APR tools can correctly fix.
\Cref{tab:cumulative} compares effectiveness in three scenarios:
\begin{enumerate*}
\item The best (highest) rank of the (first) correct fix
  among all APR tools' output patches (for the tools that can fix the bug):
  this corresponds to running all APR tools, and taking the most effective one
  for each bug.
\item The best (highest) rank of the (first) correct fix
  among \tool rankings of each APR tools' output patches \emph{individually}:
  this corresponds to running \tool on the patches produced by one APR tool
  at a time, and taking the best result for each bug.
\item The rank of the (first) correct fix
  in \tool's ranking of \emph{all} APR tools' output patches \emph{cumulatively}:
  this corresponds to taking the union of all APR tool patches, and
  asking \tool to rank it.
\end{enumerate*}

Overall, the effectiveness of \tool hardly changes
when used \emph{cumulatively} on all patches produced by multiple tools.
Furthermore, \tool's ranking remains generally better than
even the best one of all APR tools.
The only exception is for the top-1 rank:
the best APR tool in each case ranks the correct fix in the top position
in 61\% of the bugs;
\tool used individually improves this to 68\% of the bugs,
whereas \tool used cumulatively only does it for 52\% of the bugs.
In this scenario, 
\tool has to rank with perfect precision
a large number of patches,
many of which are very similar syntactically to the correct fix---several are correct, but many aren't---and
hence tend to have similar scores.
\tool still outperforms the best tools for the top-2 ranks
(88\% of bugs for \tool cumulative vs.\ 76\% for the best APR tool),
but it misses a few correct fixes in the top-1 rank.

\finding{In most cases, \tool improves the ranking of correct fixes compared to the original ranking produced by the APR tools: it ranks 8\% more correct fixes in the top-1, 27\% more in the top-3, and 11\% more in the top-10. \tool's effectiveness is largely independent of the APR tool that produced the patches, and of the category of bugs that are fixed.}

\paragraph{Ranking LLM-produced patches}
You may have noticed that our selection of APR tools in \Cref{sec:subjects} does not include any recent LLM-based technique.
As we explain in \Cref{sec:related-work},
LLM-based APR approaches often operate a closed, iterative patch generation/validation loop,
and produce at most one plausible patch per bug;
therefore, a ranking technique is not directly applicable.
Nevertheless, we would like to determine whether \tool can still be useful on the kinds of fixes that
are produced by state-of-the-art LLMs.

To this end, we conducted a small experiment with Defects4J bug \textit{Lang22}.
We prompted \texttt{GPT-4o mini} (temperature: 0.9)
20 times, with a zero-shot prompt showing the buggy code, a failing test input, and a request to
provide a patch that corrects the bug.
Out of the 20 patches produced by the LLM, 4 were correct.
We then submitted the 20 patches to \tool, which gave rank~2 to the first correct patch.\footnote{
  Interestingly, all 4 correct patches correspond to patch kind \emph{add conditional},
  whereas the incorrect patches belong to other patch kinds.
}
For comparison, the \passat{2} score~\cite{passatk} of the LLM is 37\%,
which indicates that one should expect to find a correct patch in the first two attempts
only about one third of the times.
While we cannot directly compare the inherently probabilistic output of an LLM
to \tool's deterministic output, this small experiment
suggests that an approach like \tool is not limited to traditional APR techniques
but remains applicable regardless of how the ranked patches were generated.

\subsection{RQ2: Comparison with Other Ranking Approaches}

In recent years, there has been a growing interest in approaches
to classify, assess, validate, and rank patches produced by APR systems~\cite{Yang2017BTC, Shibboleth2022, xTestCluster, ObjSimISSTA, Xin2017ITSOP, 10638611, Xiong2018IPC, Yuan2020TBEPR}.
We can roughly classify these approaches in two groups:
\begin{description}[leftmargin=10mm]
\item[Sel.] The first, larger group
  comprises approaches~\cite{Yang2017BTC, xTestCluster, 10638611, bats,  Xin2017ITSOP, Xiong2018IPC,xTestCluster} that
  perform an additional validation step---following the APR algorithm's validation---with the goal of \emph{selecting} a smaller collection of
  patches that are suggested to the user of the APR tool for their inspection.
  Most of these approaches are dynamic (i.e., test-based).

\item[Rnk.] The second, smaller group
  comprises approaches~\cite{ObjSimISSTA, Yuan2020TBEPR, Xiong2018IPC, Shibboleth2022} that
  expressly perform post-patch \emph{ranking}.
\end{description}
A fair, comprehensive comparison of \tool against several of these ranking tools
would be difficult to execute,
since each tool's experiments are not always comparable,
as they may target different sets of patches or the respective publications
may not include all necessary details.
Instead, we identify one ranking tool in each group
that:
\begin{enumerate*}
\item is recent, which improves the chances that its experiments details
  are available and are comparable to \tool's;
\item has been experimentally
  compared against other tools in the same group,
  and the experiments showed that it (generally) outperformed them.
\end{enumerate*}
Based on these criteria, we selected
xTestCluster~\cite{xTestCluster} as the benchmark for tools in group \emph{Sel.},
and Shibboleth~\cite{Shibboleth2022}
as the benchmark for tools in group \emph{Rnk.}.

\begin{figure}
	\centering
	\begin{subfigure}{.5\textwidth}
		\centering
		\begin{tabular}{rrr}
			\toprule
        \multicolumn{1}{c}{\textsc{\# clusters}}  & \multicolumn{1}{c}{\textsc{\# bugs}} & \multicolumn{1}{c}{\textsc{\% bugs}}      \\
        \midrule
        1 & 38 & 72\% \\
        2 & 12 & 23\% \\
        3 &  3 & 5\% \\
        \bottomrule
		\end{tabular}
		\caption{Number \textsc{\# bugs} and percentage \textsc{\% bugs} of
        53 Defects4J bugs (used in xTestCluster's evaluation)
        whose ranking of patches produced by \tool
      spans 1, 2, or 3 of xTestCluster's clusters.}
		\label{tab:withxTest}
    \end{subfigure}
    \hfill
	\begin{subfigure}{.45\textwidth}
		\centering	
	\begin{tabular}{lrr}
			\toprule
			\textsc{tool} & \textsc{top 1} & \textsc{top 2} \\ \midrule
			Shibboleth    & 43\%           & 66\%           \\ 
			\tool          & 53\%           & 92\%           \\ \bottomrule
		\end{tabular}
		\caption{Percentage of 66 Defects4J bugs (used in Shibboleth's evaluation)
        ranked within the top-1 or top-2 positions
      by Shibboleth and by \tool.}
		\label{tab:compare12}
	\end{subfigure}
	\caption{\tool comparison with the State-of-the-art}
	\label{fig:test}
\end{figure}

\paragraph{Comparison with xTestCluster}
xTestCluster~\cite{xTestCluster} is a technique
that groups APR patches in clusters
based on how similar their runtime behavior is on the available tests.
Since xTestCluster (like other techniques in group \emph{Sel.})
does not actually rank patches,
we cannot directly compare its effectiveness against \tool's.
Instead, we use xTestCluster's \emph{semantic}
clustering capabilities to assess
whether \tool's ranking---which uses mainly syntactic features---%
is also consistent with the patches' semantic behavior.
To this end, we consider all 413 patches for 53 Defects4J bugs
used in xTestCluster's experiments.
For each bug,
xTestCluster partitions its available plausible patches
into clusters according to semantic similarity:
xTestCluster produced 158 clusters in total,
3 clusters per bug on average,
from a minimum of 1 cluster to up to 15 clusters per bug.
We first feed the patches for each of these bugs
to \tool as usual;
this produces
a ranking of the patches among each bug's plausible patches.
For each bug $b$,
we identify the rank $r_b$ of the first
correct patch among those for $b$;
finally, we count how many clusters $c_b$
the patches in all ranks $1..r_b$
span.
The smaller $c_b$ is,
the more \tool's ranking is consistent with semantic similarity,
because all patches that are likely to be correct according to \tool's output
are also behaviorally similar.
Indeed, \Cref{tab:withxTest}
shows that $c_b$ is never larger than 3;
in 72\% of the cases, $c_b$ is just 1.
This suggests that \tool's ranking
is usually consistent with semantic similarity of patches.\footnote{
  Another way to look at these results is that \tool's ranking
  suggests an order in which to inspect xTestCluster's clusters
  that is likely to give better ranking to clusters with correct patches.
  This observation could be useful to choose the order in which to inspect
  xTestCluster's clusters.
}

\paragraph{Comparison with Shibboleth}
Since Shibboleth~\cite{Shibboleth2022} is applicable to ranking several different patches
produced by various APR techniques for the same bug,
it is directly comparable to \tool.
Consistently with the rest of our evaluation,
we only consider the 66 Defects4J bugs used in Shibboleth's
evaluation that include at least four plausible patches each
(for a total of 827 patches).
\Cref{tab:compare12}
shows the percentages of these bugs
where Shibboleth or \tool
ranked a correct patch in the top-1 or top-2 position.
Since Shibboleth was evaluated based on the top-1 and top-2 ranks,
we do not consider lower ranks in this comparison.
\tool generally ranks more bugs in high positions,
which suggests that \tool
is an effective technique that improves over the state-of-the-art
of patch ranking.

\finding{
  Even though it is based on syntactic features,
  \tool's ranking is usually
  consistent with clustering of patches by semantic
  similarity.
  \tool's ranking accuracy also compares favorably to the capabilities
  of the state-of-the-art ranking tool Shibboleth.
}

\subsection{RQ3: Robustness}
\label{sec:rq2}

We assessed \tool's robustness to changes in the quantity and variety
of training data with two experiments:

\begin{itemize}
\item \emph{Patch variety:}
  As explained in \cref{sec:classify-patch},
  \tool crucially relies on a syntactic \emph{patch classification},
  which assigns any given patch $p$ to a \emph{kind} $\patchk[p]$
  based on its syntactic features.
  The set of possible kinds is determined by the data used for training,
  which only includes patches modifying at most 5 lines of code.
  Below, we investigate whether the restriction to patches of this size
  affects the variety and distribution of patch kinds
  that are available for classification.
  
\item \emph{Training data:}
  Like every data-driven technique, \tool's capabilities
  depend on the size of its training data. Below, we investigate
  how \tool's effectiveness varies as it is trained
  on fewer data.
\end{itemize}

\begin{table}
	\centering
	\begin{tabular}{r r r}
     \toprule
     \multicolumn{1}{c}{\textsc{loc of patch}} & \multicolumn{1}{c}{\textsc{commits}} & \multicolumn{1}{c}{\textsc{selected}} \\
     \midrule
      $<$ \numprint{6}  & \numprint{224965} & \numprint{6583} \\
      $<$ \numprint{20} & \numprint{517416} & \numprint{8202} \\
     \bottomrule
	\end{tabular}
	\caption{How the number of \textsc{selected} bug-fixing \textsc{commits} changes if
     we allow all patches of up to 19 lines of code (bottom row)
     instead of 5 lines of code
     (top row) as done in the rest of the paper.}
	\label{tab:patch_stats_relaxed}
\end{table}

\paragraph{Patch Variety}
\Cref{tab:patch_stats_relaxed}
shows how many historical patches we found
in the same 81 open-source projects described in \Cref{sec:training-data}
if we consider all patches that modify up to 19 lines of code%
---increasing the bound of 5 lines of code used in the rest of the paper's
experiments.
We still did not include \emph{all} available commits
for two reasons:
first, in order to keep the manual effort feasible;
second, because
large-scale empirical studies of program repair
suggest that it is exceedingly uncommon that
an APR system can generate longer patches~\cite{8330203,8941305, jrepairanalysis}.
Even with a limit of 20 modified lines,
\Cref{tab:patch_stats_relaxed}
clearly shows that the new selection includes substantially more commits
(2.3 times more);
out of the \numprint{517416} available commits,
\numprint{8202}
(1.2 times more than in the rest of the paper)
were confirmed as genuine fixes
targeting an overflow, null pointer, or logical bug.

The \numprint{8202} patches
were classified following the same procedure of \Cref{sec:training}.
Despite including a substantially larger number of patches,
the obtained distribution of the kinds of patches was very similar to
the one induced by the smaller selection used in the rest of the experiments.
In particular, the ten most frequently occurring patch kinds
among the \numprint{8202} patches are the same
as those in \Cref{tab:category};
and the frequencies of logical, null pointer, and overflow
errors for each kind are very similar
(precisely, the difference in the most frequent bug category frequency
is within 2.2\%).
These results indicate that considering much larger patches
does not introduce significant changes in the variety and frequency
of path kinds;
correspondingly, the behavior of \tool would also remain similar.
Notice that our a priori restriction to only three broad
categories of bugs (logical, null pointer, overflow)
helps robustness even with a much larger selection of patches.

\paragraph{Training Data}
In order to assess to what extent \tool's ranking output
changes as we \emph{train} it with increasingly smaller training data,
we retrained \tool with three
different random \emph{samples} of
the complete training data (described in~\Cref{sec:training},
which we used in the rest of the paper's experiments):
a sample size of
4000 (60\%),
2000 (30\%),
and 500 (8\%, or an order of magnitude less than the original dataset)
out of all 6583 bug commits.
Precisely, we performed \emph{stratified sampling}~\cite{ARNAB2017213}
using bug kinds (logical, overflow, null pointer)
as strata to ensure that bugs of all kinds are adequately represented in the sample.
This resulted in the following selection of commits for each sample size:\footnote{
  We repeated 3 times the sampling for each size; we report averages of all measures.
  }
\begin{center}
  \begin{tabular}{c rrr}
    \toprule
    \textsc{sample} & \textsl{logical} & \textsl{overflow} & \textsl{null} \\
    \midrule
    500 & 87 & 28 & 385 \\
    2000 & 600 & 150 & 1250 \\
    4000 & 900 & 250 & 2850 \\
    \bottomrule
  \end{tabular}
\end{center}

\begin{table*}
  \centering
  \setlength{\tabcolsep}{3pt}
  \small
 \resizebox{\textwidth}{!}{  \begin{tabular}{r *{5}{rr} | *{3}{rr}}
    \toprule
    & \multicolumn{2}{c}{\textsc{rank}}
    & \multicolumn{2}{c}{\textsc{top 1}}
    & \multicolumn{2}{c}{\textsc{top 3}}
    & \multicolumn{2}{c}{\textsc{top 5}}
    & \multicolumn{2}{c}{\textsc{top 10}}
    & \multicolumn{2}{|c}{\textsc{worse}}
    & \multicolumn{2}{c}{\textsc{same}}
    & \multicolumn{2}{c}{\textsc{better}}
    \\
    \cmidrule(lr){2-3}
    \cmidrule(lr){4-5}
    \cmidrule(lr){6-7}
    \cmidrule(lr){8-9}
    \cmidrule(lr){10-11}
    \cmidrule(lr){12-13}
    \cmidrule(lr){14-15}
    \cmidrule(lr){16-17}
    \multicolumn{1}{c}{\textsc{sample}}
    & \multicolumn{1}{c}{\textsc{a}}
    & \multicolumn{1}{c}{\textsc{m}}
    & \multicolumn{1}{c}{\textsc{\#}}
    & \multicolumn{1}{c}{\textsc{\%}}
    & \multicolumn{1}{c}{\textsc{\#}}
    & \multicolumn{1}{c}{\textsc{\%}}
    & \multicolumn{1}{c}{\textsc{\#}}
    & \multicolumn{1}{c}{\textsc{\%}}
    & \multicolumn{1}{c}{\textsc{\#}}
    & \multicolumn{1}{c}{\textsc{\%}}
    & \multicolumn{1}{|c}{\textsc{\#}}
    & \multicolumn{1}{c}{\textsc{\%}}
    & \multicolumn{1}{c}{\textsc{\#}}
    & \multicolumn{1}{c}{\textsc{\%}}
    & \multicolumn{1}{c}{\textsc{\#}}
    & \multicolumn{1}{c}{\textsc{\%}}
    \\
    \midrule
    500 & 6.4 & 3 & 81 & 23.7 {\small \%} & 209 & 61.1 {\small \%} & 267 & 78.1 {\small \%} & 303 & 88.6 {\small \%} & 131 & 38.3 {\small \%} & 198 & 57.9 {\small \%} & 13 & 3.8 {\small \%}
\\
2000 & 5.7 & 2 & 102 & 30.1 {\small \%} & 230 & 67.8 {\small \%} & 283 & 83.5 {\small \%} & 310 & 91.4 {\small \%} & 90 & 26.5 {\small \%} & 230 & 67.8 {\small \%} & 19 & 5.6 {\small \%}
\\
4000 & 5.1 & 2 & 105 & 30.4 {\small \%} & 250 & 72.5 {\small \%} & 295 & 85.5 {\small \%} & 318 & 92.2 {\small \%} & 76 & 22.0 {\small \%} & 248 & 71.9 {\small \%} & 21 & 6.1 {\small \%}
\\
\midrule
\multicolumn{1}{c}{\textsl{all}} & 4.1 & 2 & 121 & 35.4 {\small \%} & 266 & 77.8 {\small \%} & 300 & 87.7 {\small \%} & 322 & 94.2 {\small \%} &  &  &  &  &  & 
    \\
    \bottomrule
  \end{tabular}}
  \caption{Robustness of \tool with training data of different size.
    The table shows how the \textsc{a}verage (mean) and \textsc{m}edian rank,
    and the number \textsc{\#} and percentage \textsc{\%}
    of bugs whose correct patch (produced by any tools)
    is ranked in the \textsc{top $k$}
    vary as \tool is \emph{trained} with a different \textsc{sample}
    of the training data: 500 commits, 2000 commits, 4000 commits,
    and \textsl{all} 6583 commits.
  In the right-hand side, it shows the number \textsc{\#} and percentage \textsc{\%}
    of bugs whose correct patch is ranked \textsc{worse}, the \textsc{same}, and \textsc{better} when \tool uses the sampled commits as training data compared to when it uses all commits.}
	\label{tab:robustness}
\end{table*}

After retraining \tool using each sample, we compared its
ranking results (on the same APR patches used in the rest of the evaluation)
to those obtained in RQ1 (i.e., with \tool trained on all 6583 bug commits).
\Cref{fig:scatterplot-robustness} and 
\Cref{tab:robustness}
show the results of these experiments, comparing them to the baseline
(bottom row, corresponding to \tool trained with all available data).
As one should expect,
the effectiveness of \tool degrades when it is trained with
less data:
this applied to pretty much all measures, from the average rank
of a correct fix to the number/percentage of bugs whose correct
fix is ranked in the top-1, top-3, top-5, or top-10.
Nevertheless, the worsening of ranking accuracy is gradual
and not dramatic:
the average rank of a correct fix, for example,
worsens by only 2.3 positions (from 4.1 to 6.4)
when reducing the training data to only 500 samples.
Similarly, the number of correct fixes ranked in the top-10
decreases by only 4 (from 322 to 318, corresponding to 2 percentage points)
when training on 4000 samples.
In all cases, the ranks of the majority of bugs do not change
(columns \textsc{same} in \Cref{tab:robustness}),
whereas the ranks of between $1/5$ and $2/5$ of all bugs worsen
(columns \textsc{worse}).
Interestingly, for a few bugs, the ranking improves
when training \tool on a smaller sample (columns \textsc{better}).
Like every statistical approach that learns from data,
\tool's behavior is not deterministic and possibly inconsistent;
importantly, our experiments show that such anomalies
occur only sporadically and do not qualitatively affect
\tool's overall capabilities.
The granularity of patch features (used to classify patches into kinds, as explained in \Cref{sec:classify-patch}) underlies \tool's robust behavior:
there still is a variety of features
even if the training data is curtailed,
which is often sufficient to still obtain a good-quality ranking.

\finding{%
  \tool's effectiveness degrades gracefully as the size of the training data shrinks:
  even when less than 8\% ($500/6583$) of the training data is used,
  the rank assigned to 57.9\% of the bugs does not change.
}

\iflong
\begin{figure}
  \centering
  \includegraphics[width=\textwidth]{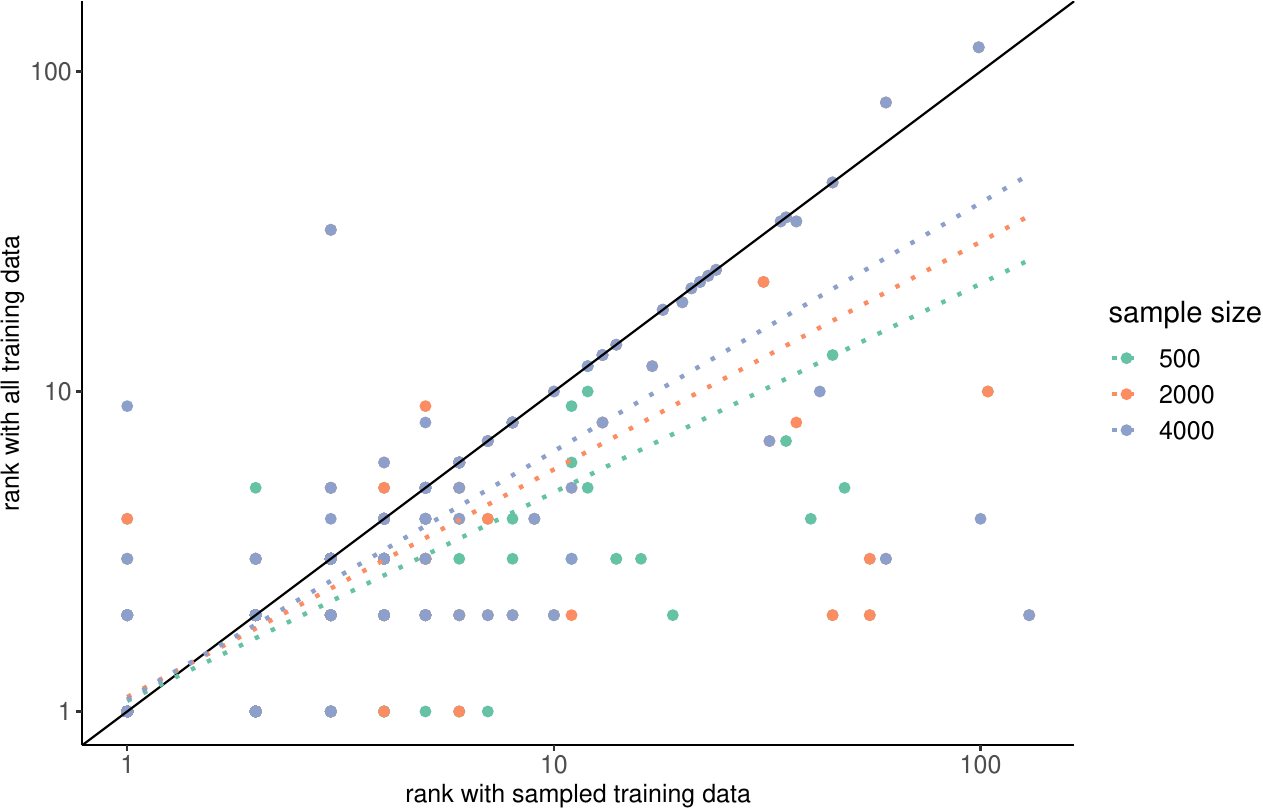}
  \caption{Each point represents a bug fixed by an APR tool. The point's $x$ coordinate is the rank assigned by \tool when it uses a randomly sampled subset of the training data; the point's $y$ coordinate is the rank assigned by \tool when it uses all training data.
    Thus, points below the diagonal line correspond to bugs where using all training data improves the ranking over using only a subset. The dotted lines are the linear regression lines of the points in each group. Axis scales are logarithmic.}
  \label{fig:scatterplot-robustness}
\end{figure}
\fi

\subsection{RQ4: Automation of Training}

As described in \Cref{sec:classify-bug} and \Cref{sec:training},
our experiments with \tool involved a partially manual validation of training data,
aimed at ensuring that the selection of bugs and classification of programmer-written patches used to train \tool is as accurate as possible.
In this section, we describe some exploratory experiments
that assess whether this manual effort can be reduced.

\begin{table}
  \centering
  \footnotesize
  \setlength{\tabcolsep}{4pt}
	\begin{tabular}{cc r rr}
     \toprule
     \multicolumn{2}{c}{\textsc{task}}
     & \multicolumn{1}{c}{\textsc{misclassifications}}
     & \multicolumn{1}{c}{\textsc{false positives}}
     & \multicolumn{1}{c}{\textsc{false negatives}} \\
     \midrule
     $T_C$ & bug classification & 5.5\% \\
     $T_+, T_-$ & bug-fix identification & 24.7\% & 40.0\% & 9.5\% \\
     \bottomrule
	\end{tabular}
	\caption{Comparison between an LLM and the manual classification of training data.}
	\label{tab:LLM_accuracy}
\end{table}

In the time since we first developed \tool,
LLMs (Large Language Models) have made spectacular advances,
and have been applied to automate various
software engineering tasks~\cite{hou_large_2024,gu_challenges_2025}.
To determine whether they can replicate
the manual validation we performed to select \tool's training data,
we set up three tasks:
\begin{description}
\item[$T_-$] We sampled randomly
  200 commits\footnote{
    This sample size is sufficient to 
    estimate the accuracy of a binary classification
    with up to 5.7\% error
    and 90\% probability with the most conservative (i.e., 50\%)
    a priori assumption~\cite{ss-estimate}.
  }
  out of the 6583
  commits that we selected as genuine bug fixes and used for training.
  For each commit, we asked the LLM to determine whether
  it represents a genuine fix of a logical, null pointer, or overflow bug.

\item[$T_+$] We sampled randomly
  200 commits out of the 1201 ($= 7784 - 6583$)
  commits whose message matches one of \Cref{code:regex}'s regexes
  but that manual analysis discarded as \emph{not} valid bug fixes.
  For each commit, we asked the LLM to determine whether it
  represents a genuine fix of a logical, null pointer, or overflow
  bug.

\item[$T_C$] We sampled randomly
  200 commits out of the 6583
  commits that we selected as genuine bug fixes and used for training.
  For each commit, we asked the LLM to determine whether the bug
  that the commit fixes is a logical, null pointer, or overflow bug.
\end{description}

All experiments were conducted using \texttt{GPT-4o mini}.
In each query, we prompted the LLM through its API
with a commit diff and message,
the list of available bug categories
(logical, null pointer, overflow);
we asked it to answer the query in JSON format
and to provide a short reason for its output.
In taking stock of the results, we used our manual classification
as ground truth.

\Cref{tab:LLM_accuracy}
summarizes the performance of the LLM on these three tasks.
In task $T_C$, the LLM's classification of 11/200 (5.5\%) bug-fixing
commits disagrees with ours; most of the misclassifications
involve overflow errors, which the LLM classified as logical
errors (possibly because the fixes usually involve
changing a Boolean condition).
In task $T_-$, the LLM tagged $19/200$ (9.5\%) commits
as non-bug fixes,
in contrast to our manual assessment
that judged all these commits as actual bug fixes.
In task $T_+$, the LLM tagged $80/200$ (40.0\%) commits
as real bug fixes, 
in contrast to our manual assessment
that judged all these commits as non bug fixes.
Overall, the LLM misidentified $(80+19)/(200 + 200)$ (24.7\%) commits.
Given this substantial disagreement,
we revised again our manual assessment and confirmed it as accurate.
In many cases, it seems that the LLM overlooked 
developer-written commit messages that explicitly marked
a commit as addressing a ``typo'' or a ``copy-paste error'',
and focused exclusively on the patch's code,
classifying the commits
as fixing non-existent logic or null pointer bugs.

These results suggest that 
LLMs still lag behind human expertise
when it comes to precisely identifying bug-fixing commits.
Given that the manual effort involved in curating the training data for \tool
(which we estimated to be around two person-days)
can be amortized over many usages,
it is arguably worth the additional accuracy that it achieves%
---at least until LLM technology catches up.
If \tool were used on a continuous basis,
it could also be updated with additional training data
by simply supplying any new bug fix that is validated
during standard code reviews of a project,
without need for a dedicated batch manual analysis.

\finding{%
  In our experiments, state-of-the-art LLMs
  incorrectly identified around 25\% of the bug-fixing commits
  used for \tool's training phase.
}

\subsection{Threats to Validity}
\iflong
  This section addresses the main potential threats to our experiments’ internal and external validity and mitigation.
\fi

\paragraph{Internal validity}
The choice of training data and of experimental subjects may affect our
experiments' internal validity.
The training data (\Cref{sec:training})
consisted of the commit history of 81 GitHub projects.
We applied standard selection criteria (size, commits, stars)
to ensure that we only used realistic projects of good quality;
to avoid data leakage, we excluded any projects
that feature in the Defects4J benchmark collection
(which we used to evaluate \tool's ranking effectiveness).
In order to minimize the chances that the training data is noisy,
two authors performed a manual validation
to confirm that we only retained bug-fixing commits
with a precise classification of their bugs.
Despite these precautions, we cannot exclude that a
few misclassified bug-fixing commits remained in the training data.
Even if this is the case, our experiments remain valid to empirically assess
\tool's ranking effectiveness.

Limiting the patches used for training to 5 lines of code
is another possible threat to internal validity.
In our robustness experiments,
we relaxed this constraint and found that including much larger
patches (up to 20 lines of code) does not seem to bring major
changes in the frequency distributions that are learned by \tool;
this helps mitigate this threat by providing an assessment of its impact.

The robustness experiments in \Cref{sec:rq2}
also assess how \tool's performance worsens
as the training data shrink.
While the performance degradation we observed in these experiments
was gradual, we cannot rule out that
using a very different sampling strategy
(in particular, one that produces a very biased dataset)
may lead to different results.
Like all data-based approaches,
\tool's capabilities ultimately depend on the quality of the training data;
but our experiments exhibited a reasonable degree of robustness.

The selection of APR patches (the experimental subjects)
included all the patches available from a wide selection of APR tools
that have been evaluated on the Defects4J curated collection;
we only excluded subjects where ranking would make little sense
(because there are only a few patches, or they all are nearly identical):
this is a conservative choice, as including these scenarios
(where any ranking is acceptable)
could have overestimated \tool's capabilities.
We also directly used patches as they were made available
by the authors of the APR techniques that produced them;
this reduces the risk that we misuse an APR tool,
leading it to produce patches that are not representative
of its intended usage.
Finally, our analysis of experimental results
used widely accepted metrics (e.g., top-$k$, average rank)
and standard visualizations and statistics%
---which mitigates further potential threats to internal validity.

\paragraph{External validity}
concerns the generalizability of results.
We tried to evaluate \tool's effectiveness with as varied a selection
of APR tools as possible, and we noticed that its performance does not
change dramatically on different tools;
this is consistent with the fact that \tool was designed to be tool-agnostic.
At the same time, \tool is practically useful only when
used in combination with APR tools capable of producing several different
plausible patches for the same bug.
On the one hand,
several cutting-edge LLM-based APR systems do not meet this requirement,
which limits \tool's general applicability.
On the other hand,
as discussed in \Cref{sec:related-work},
there remain scenarios where more traditional APR approaches
are preferable, which is precisely where \tool can be proficiently applied.

While we did not collect evidence to claim further generalizability
(to other categories bugs, or to other programming languages),
it is possible that---thanks to their simplicity---some of \tool's basic ideas
are applicable also in different scenarios.
Properly evaluating whether this is the case requires
additional experiments and belongs to future work.

\section{Related Work}
\label{sec:related-work}

Automated program repair (APR) aspires to improve software quality and reduce software development and maintenance costs by automatically suggesting fixes to program bugs~\cite{cacm,Gazzola2019survey}. 
The most common APR techniques that target general classes of software bugs are \emph{test-driven}:
passing tests define expected (correct) program behavior
that a fix should preserve,
whereas
failing tests define incorrect behavior that a fix should prevent or rectify.
\iflong
Patches that pass all tests are referred to as
\emph{valid} or \emph{plausible};
since any collection of tests cannot completely specify program correctness,
a plausible patch passing all the given tests
can still violate other, unspecified constraints (or, more generally, fail to comply with the programmer's expectations) and, therefore, be incorrect.
\fi
Plausible but incorrect patches are said to overfit the available tests. The overfitting problem has been widely observed in test-driven APR~\cite{pei_automated_2014,monperrus_critical_2014}; several studies have further investigated its causes and characteristics~\cite{qi_analysis_2015,Mechtaev2018TAA}. Despite these efforts, overfitting remains a major challenge for current repair systems~\cite{DinhLe19rpc,Wang2021apca}.

Test-driven APR techniques can be loosely classified into constraint-based, heuristic, and learning-based, depending on how they synthesize candidate patches.
\emph{Constraint-based} techniques replace suspicious expressions in the faulty program with symbolic variables (``holes''), 
build constraints on those variables w.r.t.\ all the available tests, 
and then solve the constraints to find valid fixes that are plausible by construction~\cite{avgustinov2015,xuan2016,mechtaev2016, Sergey18spr}.
\iflong
Since a constraint must encode program behavior in purely logical form,
practical challenges that constraint-based techniques face include handling expressions with side effects and generating patches that introduce complete statements.
\fi

At a high level,
all constraint-based techniques search over a space of candidate patches for plausible ones.
\iflong
  The search space is defined by the program locations that may be faulty and by the modification patterns applicable to those faulty locations (the so-called ``fix ingredients''~\cite{fix-ingredients}), and validation is done by applying each candidate patch to the program and checking whether it passes all tests.
\fi
The majority of test-driven APR techniques use heuristics to guide the search. Early examples include GenProg~\cite{weimer2009} and PAR~\cite{kim_automatic_2013}. Other approaches, such as SPR~\cite{long_staged_2015} and history-driven repair~\cite{le_history_2016}, employ novel search-based strategies based on exploiting different kinds of information. Techniques such as ssFix~\cite{Xin2017}, Jaid~\cite{pei_jaid_2017}, CapGen~\cite{WenCWHC18}, and Hercules~\cite{hercules} make the search for a patch more efficient
by abstracting the search space and selecting patch ingredients. Other systems explored similar ideas but along novel directions~\cite{pei_automated_2014,Xiong_2017,HuaZWK18,prapr}.

\emph{Learning-based} approaches inductively learn patterns programmer-written fixes in project histories, and use the learned information to improve (parts of) the APR process. Early work focused on learning repair patterns to guide template instantiation or transformation selection, thus prioritizing the generation of fixes that resemble human-written patches in the training set~\cite{long_genesis_2017,jiang2018shaping}. Other approaches trained probabilistic models to estimate whether a plausible patch is likely to be correct, which supports patch candidate ranking or filtering~\cite{honsel2015,DeepRepair,Liana2023}.

With the rise of deep learning, multiple approaches have framed program repair as a neural machine translation problem, leveraging sequence-to-sequence architectures to transform buggy code into repaired code~\cite{Ratchet,tufano2019eslbf,chen2019sequencer}. Subsequent research incorporated syntactic structure into neural models, for example by leveraging AST representations, grammar-constrained decoding, or structured edit generation to improve syntactic validity and patch quality~\cite{Codit,lutellier2020coconut,zhu2021syntax}. More recent systems enhance neural program repair by incorporating richer contextual modeling, structured decoding mechanisms, iterative refinement strategies, or execution-based feedback to improve patch correctness and handle more complex bugs.~\cite{li2020dlfix,ye2022neural,jiang2023knod,ye2024iter}.

A recent survey~\cite{LbAPR2023} systematically reviews a comprehensive selection of learning-based APR techniques. The survey presents the general idea of learning-based APR as a workflow comprising several stages. These stages include fault localization, data representation, patch generation, patch ranking, patch validation, and correctness assessment. This pipeline view suggests that each repair technique can focus on different stages of the overall process. In fact, many learning-based and LLM-based systems mainly focus on generating and validating patches; they often achieve this through repeated generate-then-validate cycles. In contrast, explicit ranking methods aim to improve the order of candidate patches that have already been made. The present paper's \tool fits into this ranking stage. It works on candidate patches after they are generated and prioritizes those that are more likely to be correct. As a result, it is independent of patch-generation methods and can be used with various repair approaches.

Like nearly every other field in software engineering, automated program repair has been influenced by Large Language Models (LLMs).  Early LLM-based APR studies explored zero-shot repair settings and prompt-based generation~\cite{xia2022less}. 
Subsequent work integrated LLMs into copilot-style or completion-based repair pipelines~\cite{wei2023copiloting}.  Other approaches investigated entropy-based scoring to assess patch plausibility~\cite{Xia2023ELLM}.  Fine-tuning strategies tailored large language models specifically for APR tasks have also been proposed~\cite{Huang2023ESFTLLM}.  Template-guided and hybrid repair methods further extended LLM-assisted APR~\cite{Gamma2023,TGPR2025}. More recent systems adopt conversational, agent-based, or self-directed paradigms to iteratively generate and refine patches~\cite{ChatRepair2024,ThinkRepair2024,RepairAgent2025,PReMM2025}. Self-directed repair strategies have also been explored~\cite{ThinkRepair2024}.

Early LLM-based APR techniques often rank potential fixes by metrics like entropy~\cite{PGwLM2022, xia2022less, Xia2023ELLM}—where a lower score suggests a more ``natural''~\cite{naturalness2012} patch. 
However, most newer approaches forgo explicit ranking and instead rely primarily on the LLM's generative capabilities. 
Several representative systems follow this paradigm~\cite{wei2023copiloting,Huang2023ESFTLLM,Gamma2023}. 
Other recent approaches similarly adopt generate-and-validate workflows without a separate ranking component~\cite{ChatRepair2024,ThinkRepair2024,TGPR2025,RepairAgent2025,PReMM2025}.
Within this context, \tool remains a relevant contribution
in two ways.
First, as we argued in \Cref{sec:llms-limitations},
while LLMs possess impressive code capabilities, their application in APR can be limited by budget, security, or practical constraints.
Second, recent studies~\cite{TGPR2025,LLM-apr-survey}
have shown that traditional APR approaches remain complementary to LLM-guided repair, especially when the required fix is small and syntactically localized. Edits such as adding a null check, adjusting a conditional, or correcting a simple assignment or initialization are still effectively handled by template-based techniques~\cite{TGPR2025}, which avoid many of the downsides and overhead of using LLMs. While state-of-the-art
LLM-driven systems routinely demonstrate strong overall APR performance, their results often depend heavily on the quality of fault localization, the size of the underlying model, and the specific inference setup~\cite{empirical-general-LLM-APR,LLM-apr-survey}.
By helping these traditional APR approaches better rank their candidate fixes, \tool directly enhances their effectiveness and practical usability.

\iflong
The rest of this section reviews various approaches
to rank plausible patches 
so that those more likely to be correct
are generated, validated, or suggested earlier.
Note that deep-learning models naturally incorporate
information, in the form of a rich feature collection,
that can be used to predict which patches are more likely correct.
\fi

\paragraph{Repair context}
APR's redundancy assumption is the claim that
large programs contain the seeds of their own repairs;
APR techniques built on this hypothesis
prefer patches that share characteristics with the code's repair context.
For example, GenProg~\cite{weimer2009} is based on the assumption that defects can be repaired by taking code elements from other locations in the program under repair.
ssFix~\cite{Xin2017} matches contextual information at the fixing location to a database of human-written fixes, and uses the matching code elements to drive patch generation.
SimFix~\cite{jiang2018shaping} combines the information extracted from existing patches and snippets similar to the code under repair to make the search for correct fixes more efficient.
CapGen~\cite{WenCWHC18} prioritizes fix ingredients based on a notion of programming context. 
Hercules~\cite{hercules} identifies a set of repair locations with similar code and likely demanding similar repairs and builds multi-hunk fixes that modify all those locations. 

\paragraph{Historical fixes}
Given that both program bugs and their corrections are highly repetitive
(that is, new bugs are often similar to old bugs, and are fixed by similar repairs), deployed programmer-written fixes in history are considered another important source of information to help prioritize candidate patches. 
For instance, PAR~\cite{kim_automatic_2013}, Jaid~\cite{pei_jaid_2017}, Restore~\cite{restore}, and TBar~\cite{tbar} construct candidate fixes by instantiating a group of repair patterns constructed manually or summarized from the literature;
Tenure~\cite{Tenure2023} and Gamma~\cite{Gamma2023} uses a deep neural network trained on human-written fixes to predict which predefined repair patterns should be selected for repairing a new bug;
SPR~\cite{long_staged_2015} systematically generates candidate fixes according to a set of predefined transformation functions;
Prophet~\cite{honsel2015} adds on top of SPR a probabilistic model, learned from human-written fixes, to effectively prioritize the generated candidate patches.
The effectiveness of these techniques is ultimately constrained by the number and diversity of their predefined repair patterns and transformation functions.
To overcome this limitation,
more recent approaches directly learn characteristics from correct fixes
and apply the characteristics to guide program repair.
The history-driven approach~\cite{le_history_2016} is based on a stochastic search process that views patches as mutants, and prefers the mutants that match the characteristics learned from fix history.
Genesis~\cite{long_genesis_2017} infers code transformations with template variables from human-written patches.
Elixir~\cite{Saha_elixir_2017} trains a model that captures the characteristics of correct fixes of buggy method invocations,
and utilizes the trained model to guide the repair of the same type of bugs.
The BATS approach~\cite{patch-correctness} is
based on the hypothesis that
bugs that are revealed by similarly failing tests
require similar patches\iflong;
it trains a deep-learning model to recognize associations between similar tests
and similar historical repairs, and then uses the trained model
to recognize patches that are more likely to be correct\fi.
FixMiner~\cite{koyuncu2020fixminer} extracts patterns of correct fixes as sequences of edit actions on a program's abstract syntax tree.
TypeFix~\cite{TypeFix2024} automatically mines, from past fixes, a collection of templates with common edit patterns and bug context information, and invokes a code pre-trained model to instantiate the templates for repairing new bugs.

\paragraph{Dynamic information} %
Given that the tests, given as input to an APR tool,
define which patches are plausible, 
it is natural for APR techniques to prioritize candidate patches
based on their impact on test execution. 
In approaches based on genetic algorithms~\cite{arcuri_novel_2008, arcuri_evolutionary_2011, weimer2009},
the fitness function typically includes information that comes from validation:
the more tests a candidate fix passes, the ``fitter'' it is.
Restore~\cite{restore} updates the suspiciousness values of various program locations by factoring in the validation results of candidate patches generated at those locations.
\iflong
Inspired by the competent programmer hypothesis~\cite{CPH1978}, v\else V\fi arious APR techniques generate new tests \iflong for the program \fi and prefer candidate patches that have less impact on the program's existing correct behavior~\cite{Yang2017BTC, Xiong2018IPC, Yu2019APOVATG, Cashin2019UAGP, Yuan2020TBEPR, Shibboleth2022}.
Such techniques differ in how they model program behavior, and in
how they quantitatively compare different behaviors.
DiffTGen~\cite{Xin2017ITSOP} follows a different approach,
wherein it generates new tests to uncover semantic differences between the original faulty program and the repaired program;
developers can then investigate its output, which helps them
discriminate between overfitting and correct patches.

\iflong
Compared with the existing fix prioritization techniques, \tool ranks plausible fixes according to their feature similarity with historic programmer-written fixes for similar bugs.
\tool implements simple heuristics, which helps make it scalable and applicable to any APR tool that produces plausible patches.
\fi

\section{Conclusions and Future Work}
This paper presented \tool: a lightweight technique
to rank the patches produced by any APR technique.
\tool uses the frequencies of similar patches
in historical project repositories
to suggest which APR patches are more likely to be genuinely correct.
Experiments demonstrated that \tool substantially improves the
ranking of correct patches (e.g., 27\% more ranked in the top-3)
and is quite robust with respect to the size of the training data.

We envision two natural directions for future work.
First, we will consider using more sophisticated machine learning techniques
than the simple frequency-based classification currently used by \tool:
while simplicity was a deliberate design feature to ensure scalability
and generality, trading off some of these features for even better effectiveness
may be advantageous in some cases.
Second, we will investigate how \tool generalizes to rank patches
for other types of bugs or to programming languages other than Java; its simple design may help enable such generalizations.
\iflong
\fi
\section{Data Availability}
\label{sec:data-availability}
\finding{
	Our prototype implementation of \tool, as well as the detailed experimental results, 
	are available at \url{https://figshare.com/s/21a42b5f72978e803768}
	}
\clearpage
\bibliographystyle{elsarticle-num} 
\bibliography{ranking-bib.bib} 
\end{document}